\documentclass[aps,prc,reprint,nofootinbib,superscriptaddress]{revtex4-2}

\usepackage[latin1]{inputenc}
\usepackage{hyperref}
\usepackage{amsmath}
\usepackage{amsfonts}
\usepackage{amssymb}
\usepackage{graphicx}
\usepackage{color}
\usepackage[per-mode=reciprocal]{siunitx}
\usepackage{multirow}

\begin{document}

\title{Characterization of ultracold neutron production in thin solid deuterium films at the PSI Ultracold Neutron source}

\author{G.~Bison}
\author{B.~Blau}
\affiliation{Paul Scherrer Institut, CH-5232 Villigen-PSI, Switzerland}
\author{W.~Chen}
\author{P.-J.~Chiu}
\affiliation{Paul Scherrer Institut, CH-5232 Villigen-PSI, Switzerland}
\affiliation{ETH Z\"urich, Z\"urich, Switzerland}
\author{M.~Daum}
\affiliation{Paul Scherrer Institut, CH-5232 Villigen-PSI, Switzerland}
\author{C. B.~Doorenbos}
\author{N.~Hild}
\author{K.~Kirch}
\author{V.~Kletzl}
\affiliation{Paul Scherrer Institut, CH-5232 Villigen-PSI, Switzerland}
\affiliation{ETH Z\"urich, Z\"urich, Switzerland}
\author{B.~Lauss}
\email[Corresponding author: 
]{bernhard.lauss@psi.ch}
\affiliation{Paul Scherrer Institut, CH-5232 Villigen-PSI, Switzerland}
\author{D.~Pais}
\author{I.~Rien\"acker}
\email[Corresponding author: 
]{ingo.rienaecker@psi.ch}
\affiliation{Paul Scherrer Institut, CH-5232 Villigen-PSI, Switzerland}
\affiliation{ETH Z\"urich, Z\"urich, Switzerland}
\author{D.~Ries}
\affiliation{ Department of Chemistry - TRIGA site, Johannes Gutenberg University, 55128, Mainz, Germany}
\author{P.~Schmidt-Wellenburg}
\affiliation{Paul Scherrer Institut, CH-5232 Villigen-PSI, Switzerland}
\author{V.~Talanov}
\affiliation{Paul Scherrer Institut, CH-5232 Villigen-PSI, Switzerland}
\author{G.~Zsigmond}
\affiliation{Paul Scherrer Institut, CH-5232 Villigen-PSI, Switzerland}

\begin{abstract}
	We determined the ultracold neutron (UCN) production rate by superthermal conversion in the solid deuterium (sD$_2$) moderator of the UCN source at the Paul Scherrer Institute (PSI). In particular, we considered low amounts of less than 20\,\si{\mol} of D$_2$, deposited on the cooled moderator vessel surfaces in thin films of a few mm thickness. We measured the isotopic ($c\textsubscript{HD} < 0.2$\,\%) and isomeric ($c\textsubscript{para} \le 2.7$\,\%) purity of the deuterium to conclude that absorption and up-scattering at 5\,\si{\K} have a negligible effect on the UCN yield from the thin films. We compared the calculated UCN yield based on the previously measured thermal neutron flux from the heavy water thermal moderator with measurements of the UCN count rates at the beamports. We confirmed our results and thus demonstrate an absolute characterization of the UCN production and transport in the source by simulations. 
\end{abstract}

\maketitle

\section{Introduction}
\label{sec:introduction}

Ultracold neutrons (UCN) are free neutrons with kinetic energies below about 300 neV that experience total reflection under all angles of incidence from surfaces of suitable materials \cite{Zeldovich}. Thus, UCN can be contained in storage bottles for hundreds of seconds by surface reflections, as well as by magnetic and gravitational confinement \cite{Steyerl1969,Lushchikov1969,Ignatovich1990,Golub1991}. The storage of UCN is a technique well suited for precision experiments that require long observation times. A prominent example is the search for a permanent electric dipole moment of the neutron (nEDM) \cite{Altarev1980,Baker2006,Pendlebury2015}, which tests physics beyond the Standard Model \cite{Engel2013a}. A sizeable nEDM would provide a new source of CP violation, one necessary ingredient of baryogenesis \cite{Sakharov1991}, that could help to explain the matter anti-matter asymmetry in the Universe \cite{Morrissey2012}. The to-date most stringent limit on the nEDM \cite{Abel2020} was obtained at the UCN source at the Paul Scherrer Institute (PSI) \cite{Bison2020}, currently hosting the follow-up experiment n2EDM \cite{Ayres2021n2edm} aiming for a tenfold improvement in sensitivity. Achieving these high sensitivities requires high UCN statistics. Hence there is an ongoing effort to fully understand the physics of the source, to characterize all its aspects, and to identify components for possible improvements.\\

\begin{figure}[t]
	\begin{center}
		\resizebox{.45\textwidth}{!}{\includegraphics{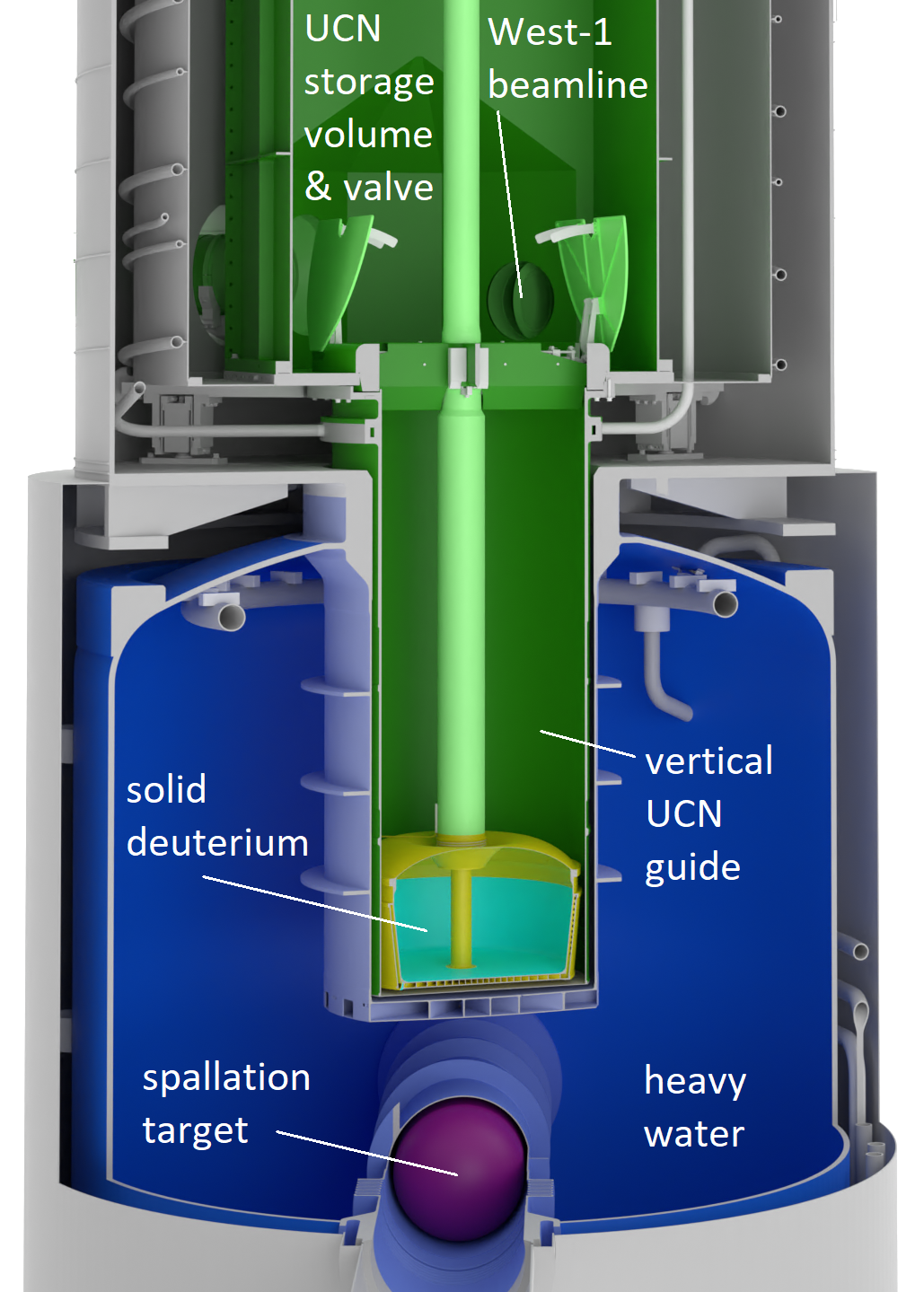}
		}
	\end{center}
	\caption{Drawing of the lead spallation target (\textbf{purple}) surrounded by the heavy water moderator (\textbf{blue}). The sD$_2$ moderator vessel (\textbf{yellow}) is suspended from above into the vertical UCN guide. The cooled surfaces at the bottom and side wall of the sD$_2$ moderator vessel are indicated in \textbf{cyan}. Closing of the valve at the top of the vertical guide confines the UCN in the storage vessel and beamlines above. The surfaces of the UCN storage and transport system are indicated in \textbf{green}, including the neutron guide shutters of the beamlines, visible in the drawing at the opening towards the West-1 beamline.}
	\label{fig:source}
\end{figure}

The PSI UCN source moderates free spallation neutrons from a lead target \cite{Wohlmuther2006} produced in up to 8\,\si{\s} long pulses of the 590\,\si{\mega \eV} proton beam of the High Intensity Proton Accelerator (HIPA) at beam currents of up to 2.4\,\si{\milli \ampere} \cite{Grillenberger2021}. The target is surrounded by a heavy water moderator thermalizing neutrons at room temperature. Thermal neutrons entering the solid deuterium (sD$_2$) moderator vessel, itself cooled by supercritical helium to 5\,\si{\K}, are further moderated to cold neutrons with energies below 10\,\si{\milli \eV} which can be converted to UCN by phonon excitations \cite{Golub1983}. The UCN accumulate in the evacuated storage volume above the moderator vessel and can be guided to the experiments along three beamlines. Figure \ref{fig:source} shows an illustration of the deuterium moderator vessel (yellow) and the UCN transport system (green). A full characterization and the simulation model of the neutron optics of the beamlines and storage volume was published recently \cite{Bison2020,Bison2022}.\\

Neutron absorption and scattering \cite{Liu2000,Morris2002,Atchison2005b} limit the extraction efficiency, i.e.\, the fraction of UCN that escape from the bulk sD$_2$ into vacuum. However, UCN losses become negligible in the case of thin films of sD$_2$ with thicknesses much smaller than the mean free path for UCNs in sD$_2$. Indeed, a thin film source for ultracold neutrons was already conceptualized in \cite{Golub1983,Yu1986}. We profit from vanishing UCN losses and vanishing moderation in thin films of sD$_2$ deposited on the cooled walls of the moderator vessel, to determine the UCN production rate and test its calculation and our source model. For this we rely on the previously simulated and measured thermal neutron flux \cite{Becker2015} from the heavy water moderator as input. The benchmarked UCN transport simulation code MCUCN \cite{Zsigmond2018} and model of the source \cite{Bison2020,Bison2022} can be used to relate the computed UCN downscattering rate to detected UCN count rates at the beamports. A verification of the UCN yield from a thin film by measurements at the beamports can be considered as a full and absolute characterization of the neutron production by spallation, thermal moderation, UCN conversion, storage, and transport in the PSI UCN source. Furthermore, together with MCNP6 \cite{Goorley2012} simulations of cold moderation in solid deuterium, it allows us to study the UCN extraction from standard sD$_2$ filling levels of 13\,\si{\centi m} \cite{Bison2020} in the moderator vessel, i.e.\, 1100\,\si{mol} of deuterium, and thus the quality of the sD$_2$ polycrystal in a next step.\\

In section \ref{sec:ThermalDownscattering} and \ref{sec:UCNLossTransport} we discuss our simulation and computation models, including the thermal neutron spectrum, the evaluation of the downscattering cross section, and the UCN transport. In addition, we present the measured high isotopic and isomeric purity of the deuterium to demonstrate their negligible contribution to UCN losses in the thin film source. The procedure to fill the moderator vessel with small, accurately known amounts of sD$_2$ with an uncertainty of approximately 0.1\,\si{\mol}, and the measurements of the corresponding UCN yield is described in section \ref{sec:sD2} and \ref{sec:measurement}. We compare these measurements with the predicted UCN yield by our simulation models in section \ref{sec:analysis} and find good agreement within model uncertainties. 

\section{Thermal neutron flux and UCN downscattering}
\label{sec:ThermalDownscattering}

\begin{figure}[b]
	\begin{center}
		\resizebox{.51\textwidth}{!}{\includegraphics{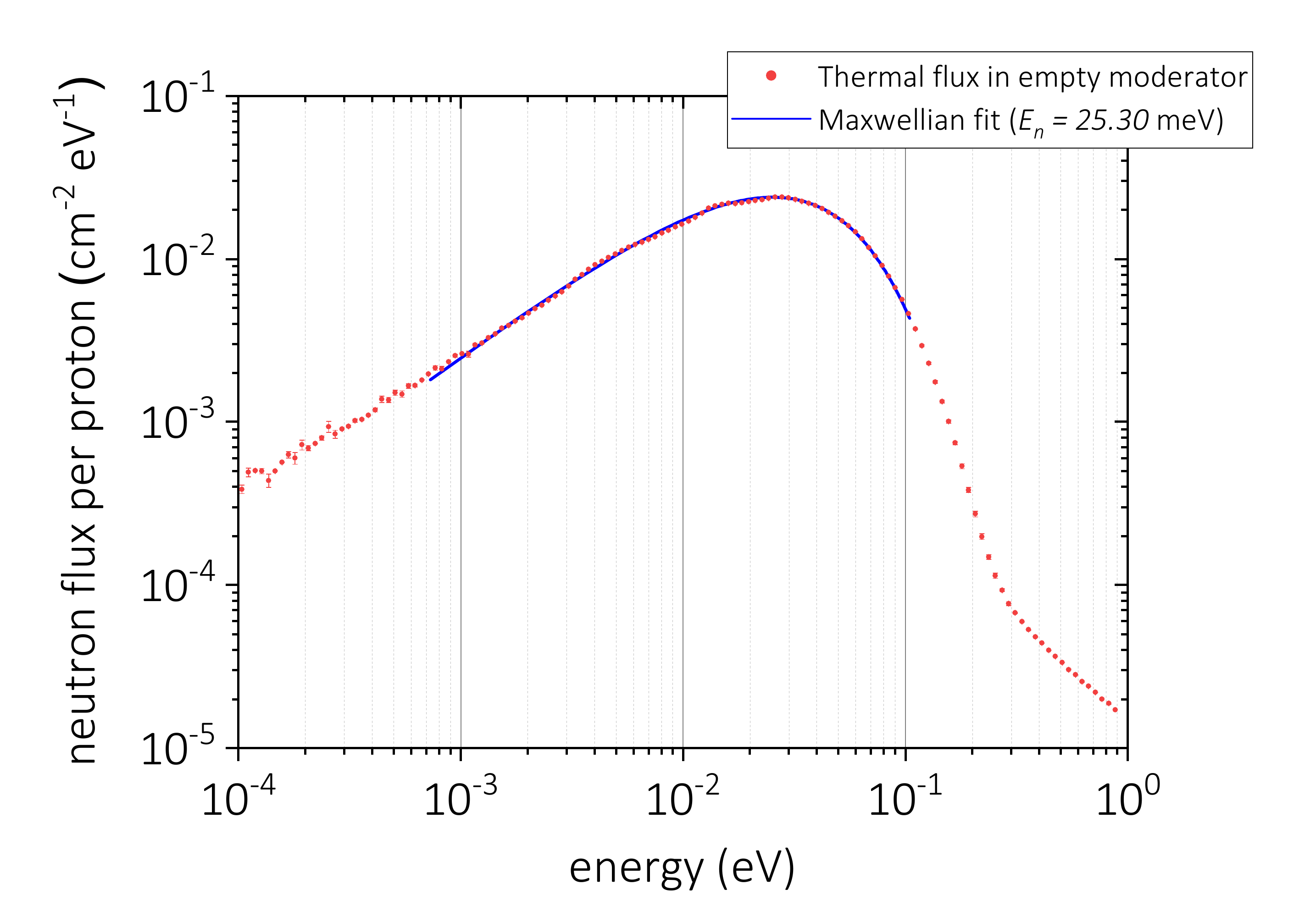}
		}
	\end{center}
	\caption{Simulated energy spectrum of the average neutron flux per primary proton in the empty moderator vessel (\textbf{red points}). The \textbf{blue line} indicates a Maxwellian fit to the thermal spectrum with $E_n=25.30\,\si{\milli \eV}$, corresponding to $T_n = 293.5$\,\si{\K}, and $\Phi_0 = 1.6 \times10^{-3}\,\si{\per \square \centi \m}$ per proton on the spallation target.}
	\label{fig:flux}
\end{figure}

A detailed geometry model of the source was implemented in MCNP for the determination of the thermal and epithermal neutron flux from the target and heavy water moderator. The measured activation of gold foils with and without cadmium shielding \cite{Becker2015} was consistent with MCNP simulations of the (epi)thermal neutron flux. We employed the identical geometrical model and performed a MCNP6 \cite{Goorley2012} simulation to tally the neutron flux in the empty moderator vessel. We confirmed that in the energy range of interest, between $10^{-3}\,\si{\eV}$ and $10^{-1}\,\si{\eV}$, the flux decreases from the bottom to the top of the empty moderator vessel by less than 15\,\%. The energy dependence of the neutron flux $\frac{d\phi}{dE}|\textsubscript{therm}$ for this energy range is very well approximated by a Maxwell-Boltzman distribution $\frac{\Phi_0 E}{(k\textsubscript{B} T_n)^2} e^{-E/k\textsubscript{B} T_n}$ with $T_n = 293.5$\,\si{\K}, as shown in Fig. \ref{fig:flux}. As the variation was less than 15\,\%, we averaged the neutron flux over the moderator vessel volume to obtain $\Phi_0 = \int_{0}^{\infty} \frac{d\phi}{dE} dE = 1.6\times10^{-3}\,\si{\per \square \centi \m}$ per proton on target.  \\

Based on a simulated neutron flux $\frac{d\phi}{dE}(E)$, we compute the specific UCN downscattering rate per unit volume and per primary proton
\begin{equation}
\frac{dR}{dE}(E) = \int_{0}^{\infty} \frac{d\Sigma\rlap{\textsuperscript{5\,K}}\textsubscript{down}}{dE}(E_0,E) \frac{d\phi}{dE_0}(E_0) \ dE_0 .
\label{eq:downscattering}
\end{equation}
The macroscopic downscattering cross section from initial to final energy $E_0\rightarrow E$ in sD$_2$ with density \mbox{$\rho=3.02\times10^{22}\,\si{\per \cubic \centi \m}$ \cite{Souers1986}} at $T=5\,\si{\K}$,
\begin{equation}
	\begin{aligned}
	\frac{\Sigma\rlap{\textsuperscript{5\,K}}\textsubscript{down}}{dE} = &\rho \, \sigma \, \frac{\tau}{2E_0} \, \frac{g(E_0-E)}{(E_0-E)(1-e^{-(E_0-E)/k\textsubscript{B}T})}\\
	&e^{-\frac{E_0+E}{\mu \tau}} \, \Big((E_0+E+\mu\tau) \, \sinh\Big(\frac{2\sqrt{E_0E}}{\mu\tau}\Big)\\
	& \qquad \quad \, -2\sqrt{E_0E}\, \cosh\Big(\frac{2\sqrt{E_0E}}{\mu\tau}\Big)\Big),
	\end{aligned}
	\label{eq:downscatteringcrosssection}
\end{equation}
and the characteristic energy
\begin{equation}
	\tau = \Bigg(\int_{0}^{\infty}\frac{1}{\epsilon}\coth\Big(\frac{\epsilon}{2 k\textsubscript{B}T}\Big) g(\epsilon) d\epsilon\Bigg)^{-1},
	\label{eq:gamma}
\end{equation}
were evaluated in references \cite{Golub1983, Yu1986, Atchison2007} based on the incoherent approximation and the Debye model for the phonon density of states in sD$_2$, as well as for a more realistic phonon spectrum with about the same result. We use the Debye spectrum	$g(\epsilon) = 3 \, (k\textsubscript{B} \Theta\textsubscript{D})^{-3} \, \epsilon^2$ for $\epsilon < k\textsubscript{B} \Theta\textsubscript{D} = 9.5\,\si{\milli \eV}$ \cite{Yu1986} and compute the cross section Eq.\,\eqref{eq:downscatteringcrosssection} with $\sigma = 4 \pi b\textsubscript{eff}^2$ based on an effective scattering length per molecule and reduced mass $\mu = \frac{m\textsubscript{mol}}{m\textsubscript{n}}=4$. The effective scattering length was calculated in \cite{Frei2009} to be $b\textsubscript{eff}^2(Q) = 4\left(b\textsubscript{coh}^2+\frac{5}{8} b\textsubscript{inc}^2\right)j_0^2(Qa_s/2)$, with the spherical Bessel function of order zero $j_0$, the momentum transfer $Q=\sqrt{\frac{2 \, m E_0}{\hbar^2}}$ (for $E\ll E_0$), and the separation between the deuterons $a_s = 0.742\,\si{\AA}$\,\cite{Nielsen1971} in the molecule. Evaluation of Eq.\,\eqref{eq:downscattering} with the average thermal neutron flux $\frac{d\phi}{dE}|\textsubscript{therm}$ per \si{\micro \coulomb} protons on the spallation target yields a downscattering rate of
\begin{equation}
\frac{dR}{dE}\Big|_\text{therm}(E)=4.0\times10^{-3}\,\si{\per \cubic \centi \m \per \micro \coulomb \per \nano \eV \tothe{3/2}}\times\sqrt{E}.
\label{eq:downscatteringEtherm}
\end{equation}

\section{UCN losses and transport}
\label{sec:UCNLossTransport}

The deuterium inventory of the PSI UCN source is about 1500\,\si{\mol} which can be stored in gaseous form at room temperature. For operation, the D$_2$ is first solidified into a condenser vessel, melted, and then transferred as a liquid to a para-to-ortho converter vessel \cite{Anghel2008,Lauss2021} filled with OXYSORB$^{\text{\textregistered}}$ \cite{Sullivan1990}. The paramagnetic centers of the chromium-oxide-based catalyzer material lead to a fast para-to-ortho conversion \cite{Bodek2004} of liquid deuterium, kept at around 21.5\,\si{\K}, to achieve a high ortho-D$_2$ concentration of 97.3\,\% \cite{Hild2019}. The D$_2$ is then transferred to the moderator vessel, where it is frozen and cooled down to 5\,\si{\K}. The ortho-D$_2$ concentration was measured by Raman spectroscopy of gas samples \cite{Hild2019} collected from the vapor of liquefied deuterium in the moderator vessel via a connection to a sample port. Below, we show that the ortho-D$_2$ concentration of 97.3\,\% is sufficient for the measurements presented here. However, within weeks of operation the measured ortho-D$_2$ concentration increases to well above 99\,\% by an accelerated conversion process \cite{Collins1991,Atchison2003, Mishima2007, Wlokka2016, Hild2019} induced by radiation from the spallation target. While the H$_2$ content is negligible and below the sensitivity of the measurement, the concentration of HD compounds in the gas sample was measured to be $(0.18 \pm 0.03)$\,\%, including a correction for a distillation effect due to the different vapor pressures of D$_2$ and HD \cite{Hild2019,Souers1986}.  \\

Based on the measured isotopic and isomeric purity, as well as the sD$_2$ temperature, we can determine the UCN lifetime in the sD$_2$ of the PSI source, similar to the computations in \cite{Morris2002}. We consider zero-phonon rotational de-excitation ($J\!=\!1 \rightarrow J\!=\!0$) upscattering with $(\Sigma\textsubscript{para}v)^{-1}=2.5$\,\si{\milli \s} on a maximum remaining para-D$_2$ concentration of 2.7\,\% and thermal one-phonon upscattering at 5\,\si{\K} of $\sigma\textsubscript{phonon} = 0.25$\,\si{\barn} (at $v=7.92\,\si{\m \per \s}$), both derived in \cite{Liu2000}
Neutron absorption cross sections $\sigma\textsubscript{abs}$ of deuterium and hydrogen were taken from \cite{Sears1992}. We employed the $1/v$ scaling of the UCN loss cross sections to obtain a velocity independent total lifetime of
\begin{align}
\tau & = \frac{1}{v} \big[c_\text{para} \ \Sigma_\text{para} + \Sigma_\text{phonon} + \Sigma_\text{abs,D$_2$} + c_\text{HD} \ \Sigma_\text{abs,HD}\big]^{-1} 	\label{eq:lifetime}\\
& = \Big[\frac{1}{56 \, \si{\milli \s}}+\frac{1}{168 \, \si{\milli \s}}+\frac{1}{146 \, \si{\milli \s}}+\frac{1}{269 \, \si{\milli \s}}\Big]^{-1} \nonumber\\[3ex]
& = 29 \, \si{\milli \s}, \nonumber
\end{align}
corresponding to a mean free path of $\lambda = 9\,\si{\centi \m}$ at $v=3.2\,\si{\m \per \s}$, the minimum transmitted neutron velocity\footnotemark[1] to the UCN source beamports due to the AlMg3 vacuum separation windows in the beamlines \cite{Bison2020}. 
\\

\footnotetext[1]{Taking into account that the energy boost when exiting the sD$_2$ is compensated by the gravitational deceleration in the vertical guide above the moderator vessel.}

From the calculated mean free path and earlier measurements of the total UCN scattering cross section \cite{Atchison2005b} in sD$_2$, we conclude that UCN losses are negligible for a total path length of a few mm in sD$_2$ thin films. Thus we can directly multiply the downscattering spectrum $\frac{dR}{dE}(E)$, Eq.\,\eqref{eq:downscattering}, with the energy-dependent transmission probability $t(E)$ for a UCN starting in the empty moderator vessel to reach the detector at beamport West-1 to obtain the number of detectable UCN,
\begin{equation}
N = \int_{0}^{\infty} \frac{dR}{dE} \ t(E) \ dE.
\label{eq:result}
\end{equation}

\begin{figure}[t]
	\begin{center}
		\resizebox{.5\textwidth}{!}{\includegraphics{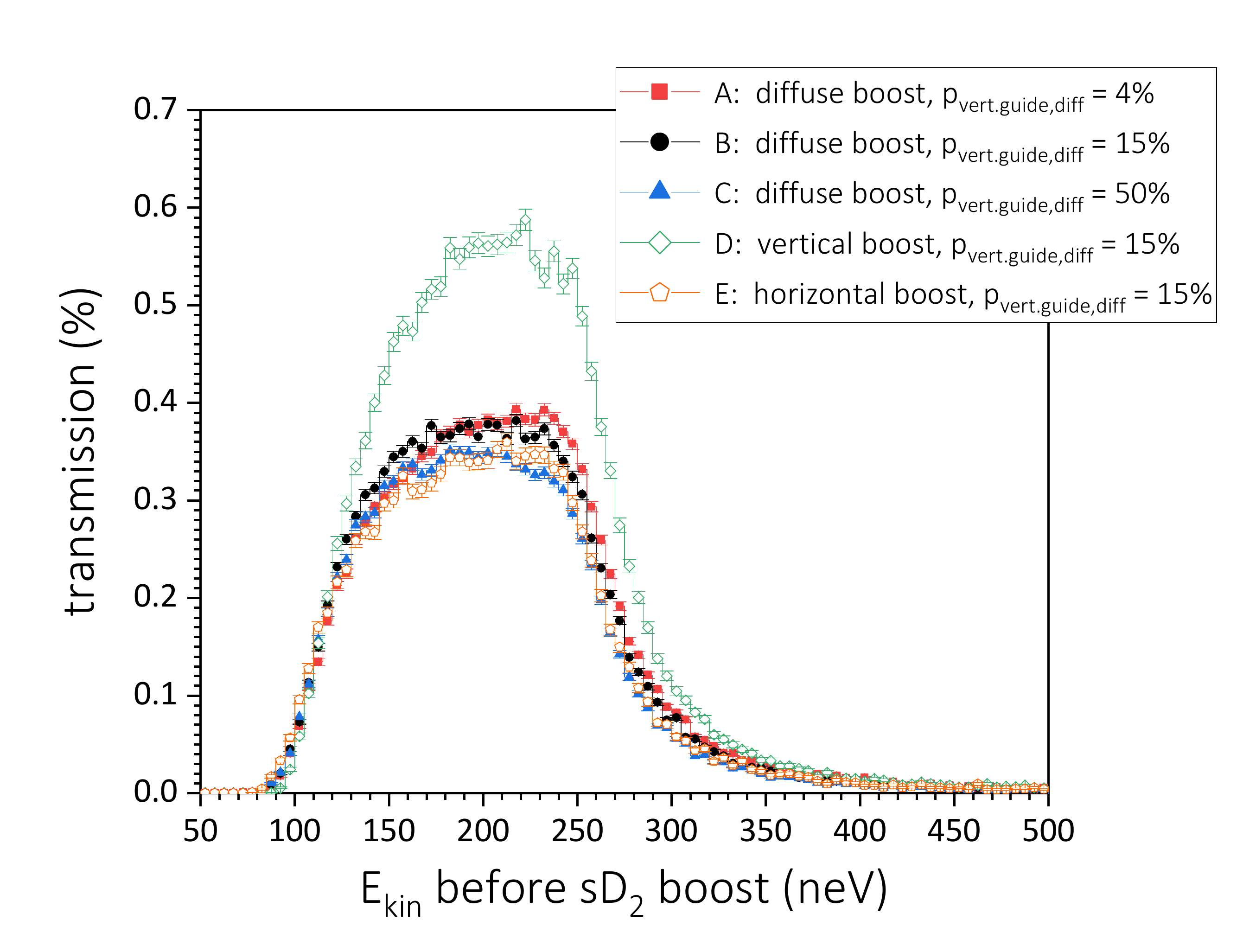}
		}
	\end{center}
	\caption{Simulated transmission probability $t(E)$ of UCN to be counted at the detector on beamport West-1 integrated over 280\,\si{\s} after generation in the moderator vessel. The \textbf{red, black and blue markers} show the transmission of configurations A, B and C of UCN starting with a diffuse boost from all cooled surfaces in the moderator vessel and different probabilities of diffuse reflection $p\textsubscript{vert.guide,diff}$ in the vertical guide below the storage volume (see Fig.\ref{fig:source}). The \textbf{green and orange open markers}, configuration D and E, show the transmission when the UCN start exclusively from the base of the moderator with a vertical boost or from the side wall with a horizontal boost, respectively. Further simulation parameters are described in the text.}
	\label{fig:transmission}
\end{figure}

The transmission probabilities shown in Fig.\,\ref{fig:transmission} as a function of the UCN kinetic energy in the moderator vessel before the sD$_2$ boost were obtained from a simulation model of the source \cite{Bison2020} with UCN optics parameters from calibration measurements \cite{Bison2022}. The distribution of starting positions and initial velocity vectors of the UCN was adjusted to reflect a production of UCN in a thin film of sD$_2$ on the moderator vessel surfaces. For configurations A, B and C shown in Fig.\,\ref{fig:transmission}, the UCNs were generated with isotropic velocities at a random position at a distance of 1\,\si{\milli \m} from the base and side wall\footnotemark[2]
\footnotetext[2]{The surface areas are 1665\,\si{\square \centi \m} for the base and 2405\,\si{\square \centi \m} of the cooled part of the side wall of the moderator vessel.}
up to a height of 16\,\si{\centi \m} inside the moderator vessel. This distribution of UCN starting positions corresponds to a uniform D$_2$ deposition from the gas phase on all surfaces in direct contact with the helium cooling channels in the wall and base of the moderator vessel \cite{Bison2020}, as discussed in more detail in section \ref{sec:sD2}. Upon leaving sD$_2$, UCN experience a velocity boost due to the Fermi potential of sD$_2$ of 105\,\si{\nano \eV}, computed based on its density at 5\,\si{\K} \cite{Souers1986} and coherent scattering length \cite{Sears1992}, confirmed within experimental uncertainty by measurements in \cite{Daum2008,Altarev2008}.  To model a high surface roughness and a potentially uneven distribution of deuterium frozen on cold surfaces in the moderator vessel \cite{Anghel2018}, we consider a diffuse boost by adding a corresponding isotropically distributed velocity component from the sD$_2$ into vacuum to the UCN starting velocity. Since the calibration measurements \cite{Bison2022} for the UCN source simulation model were not sensitive to the probability of diffuse reflection in the vertical guide below the storage vessel (see Fig.\,\ref{fig:source}), we scanned this parameter in a large range from $p\textsubscript{vert.guide,diff}=4\,$\% (configuration A) to $50\,$\% (configuration C). We also scanned other UCN optics parameters obtained from the calibration measurements within their uncertainties \cite{Bison2022} and found no significant variations of the corresponding UCN yield.\\

During the measurements and in our simulations, the UCN valve at the bottom of the storage volume stayed permanently open. The neutron guide shutters (see Fig.\,\ref{fig:source}) to the West-1 and West-2 beamlines were fully open, while the shutter to the South beamline was closed. The simulated UCNs were counted behind a 100\,\si{\micro \m} AlMg3 surface, placed at the end of a 1\,\si{\m} neutron guide mounted at beamport West-1 and an additional shutter, corresponding to the detection setup described in section \ref{sec:measurement}. The number of UCNs arriving at the virtual detector were integrated during 280\,\si{\s} after generation and divided by the number of started UCNs to obtain the transmission probability.\\

We found that for a diffuse boost, the transmission of UCN starting exclusively from the base of the moderator vessel differs only by a small amount from the transmission of UCN starting from the side walls. In contrast to that, a strictly vertical boost of UCN starting from the base of the moderator vessel (configuration D, Fig.\,\ref{fig:transmission}) leads to a significantly higher transmission and increases the UCN yield by $40\,$\% (Tab.\,\ref{tab:result}), while a horizontal boost of UCN starting from the side walls of the moderator vessel (configuration E) reduces the transmission slightly. A similar increase of approximately $40\,$\% of the measured UCN yield was observed in \cite{Ries2016} after melting small amounts of deposited sD$_2$ and re-solidifying it from the melt at the bottom of the moderator vessel. The sD$_2$ was previously deposited by a procedure identical to the one presented in section \ref{sec:sD2} in the cold moderator vessel from the gas phase, presumably as a thin film with very high surface roughness, while it was subsequently frozen from the liquid phase with a smoother surface. This observation is consistent with the hypothesis that a diffuse boost describes more accurately the velocity distribution of UCN from the deuterium deposited from the gas phase on the cold moderator vessel surfaces, i.e.\, the condition during the measurements reported here.\\

In our simulations, we also tallied the number of wall reflections on the cooled surfaces in the moderator vessel to estimate the total path length $\mathit{l}$ from multiple passages of the UCN through sD$_2$ with a maximum film thickness of 0.9\,\si{\milli \m} (see section \ref{sec:sD2}). Based on this, we confirm that the expected average attenuation $e^{-l/\lambda}$ due to neutron losses (with mean free path $\lambda = 9\,\si{\centi \m}$)  is below $2\,$\%.


\section{Preparation of solid deuterium}
\label{sec:sD2}

At the end of an annual operation cycle of the UCN source, the deuterium content of the moderator vessel was evaporated into the storage tanks \cite{Lauss2021}. During this process, some of the high-ortho-D$_2$ was collected in the condenser vessel, which was cooled below the triple point of deuterium of 18.7\,\si{\K} \cite{Souers1986}. After evaporation, the residual pressure in the 51.66\,liters\footnotemark[3] moderator vessel and auxiliary piping was 0.3\,\si{\milli \bar} at 24.5\,\si{\K}, i.e.\, less than 0.01\,\si{\mol} of D$_2$ gas. We proceeded by heating the 44.87\,liters\footnotemark[3] condenser vessel to 24.9\,\si{\K}. The quoted temperature is the average reading of two temperature sensors at the top and bottom of the vessel, which were consistent within 0.2\,\si{\K}. We obtained a stable pressure of $(858\pm2)\,\si{\milli \bar}$, corresponding to $(18.5\pm0.1)\,\si{\mol}$ of deuterium gas.\\

\begin{table}[b]
	\centering
	\begin{tabular}{m{2.3cm} m{1.8cm}| m{1.3cm} m{1.3cm} m{1.2cm}}
		sD$_2$ amounts & uncertainty & \multicolumn{3}{c}{UCN} \\
		(\si{\mol}) & (\si{\mol}) & \multicolumn{3}{c}{($10^3 \,$/\,280\,\si{\s})} \\[1pt] \hline
		& & & & \\[-0.8em]
		0.0 & 0.0 & \ 3.05(6) & 2.93(5) & 2.91(5) \\[2pt]
		0.6 & 0.1 & \ 4.70(7) & 4.77(7) & 4.89(7) \\[2pt]
		1.4 & 0.1 & \ 7.15(9) & 7.29(9) & 7.33(9) \\[2pt]
		2.4 & 0.1 & \ 10.0(1) & 10.1(1) & 10.1(1) \\[2pt]
		4.9 & 0.1 & \ 17.4(1) & 17.7(1) & 17.6(1) \\[2pt]
		9.9 & 0.1 & \ 30.7(2) & 30.8(2) & 30.6(2) \\[2pt]
		14.8 & 0.1 & \ 46.3(2) & 45.6(2) & 45.5(2) \\[2pt]
		18.5 & 0.1 & \ 58.9(2) & 58.9(2) & 58.5(2) \\[2pt]
	\end{tabular}
	\caption{Measured UCN count rates during 280\,\si{\s} after 2\,\si{\s} norm pulses at beamport West-1 for different amounts of sD$_2$. For each sD$_2$ amount three proton beam pulses were performed. The beam current was monitored during each pulse and the count rates normalized to a current of 2.2\,\si{\milli \ampere}. The quoted uncertainties on the count rates are the Poisson error.} 
	\label{tab:measurement:results}
\end{table}

To fill the moderator vessel with accurate amounts of deuterium, a valve between condenser and moderator vessel was opened, while monitoring the decrease of pressure in the condenser. The AlMg3 top lid of the moderator vessel was heated to 24\,\si{\K} (measured with a thermocouple) by thermoelectric heaters \cite{Anghel2018}, while the side wall and base of the vessel were cooled to approximately 5\,\si{\K}, indicated by the measured helium in- and outflow temperatures. We assume a deposition of deuterium on all surfaces in direct contact with the helium cooling channels on the bottom and side wall of the moderator vessel, while a deposition on the upper lid of the vessel is excluded by its temperature.  After reaching a certain pressure reduction in the condenser, the valve was closed and the UCN count rate at the beamport was measured with the known amount of sD$_2$ (see Tab.\,\ref{tab:measurement:results}). During the UCN measurements the heating of the lid was turned off.\\

For subsequent measurements with larger amounts of sD$_2$, the above described sequence was repeated and deuterium gas added to the cold moderator vessel. This way, the moderator vessel was filled with small amounts of sD$_2$, ranging from 0.6 to 18.5\,\si{\mol} with an uncertainty of 0.1\,\si{\mol}. The maximum amount corresponds to a solid thin film thickness of less than 0.9\,\si{\milli \m} if the sD$_2$ is deposited uniformly on all helium-cooled surfaces, or 2.2\,\si{\milli \m} if the sD$_2$ would be distributed uniformly only on the bottom. 

\section{UCN detection setup and measurements}
\label{sec:measurement}

\begin{figure}[t]
	\begin{center}
		\resizebox{.52\textwidth}{!}{\includegraphics{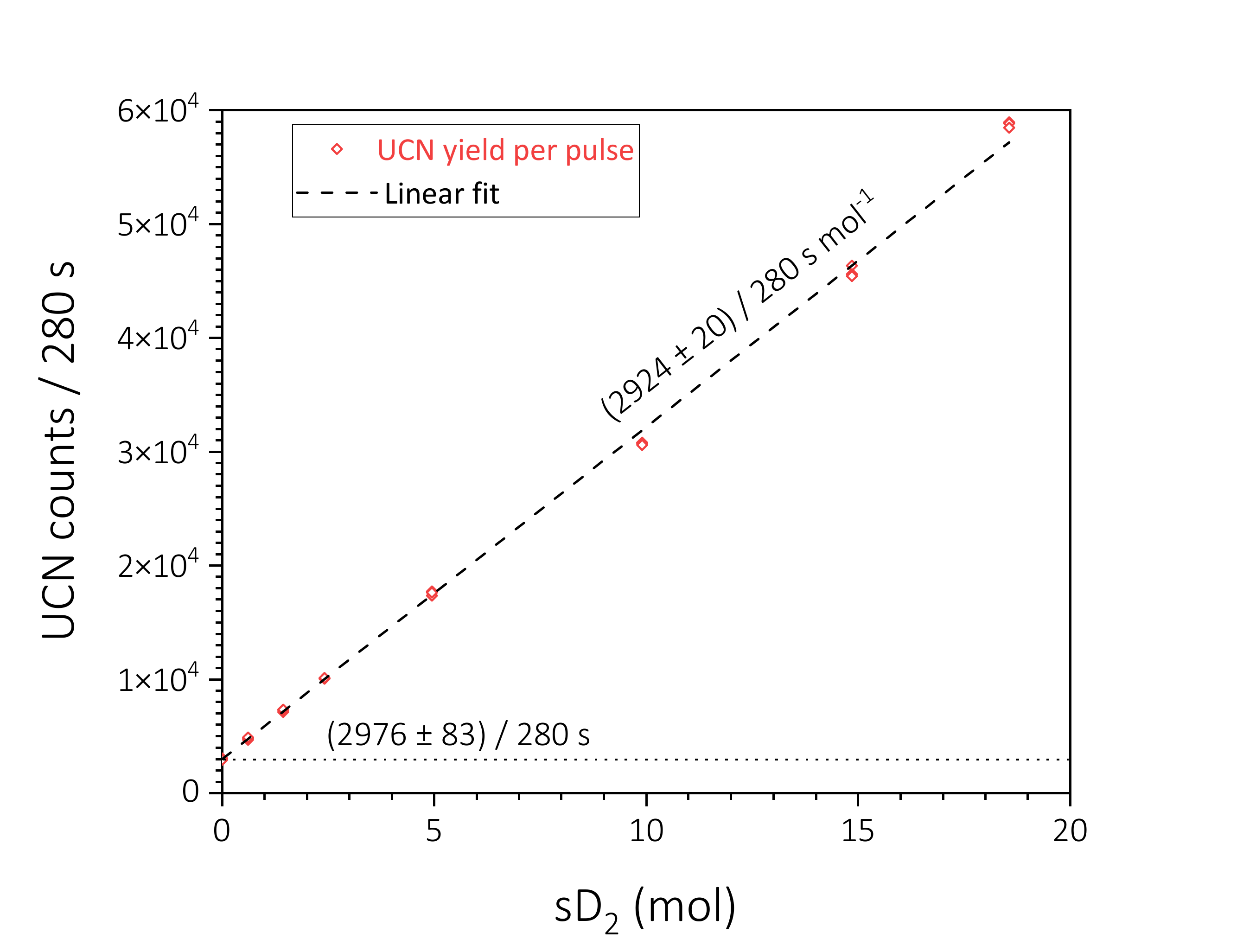}
		}
	\end{center}
	\caption{Measured UCN counts (\textbf{red squares}) at beamport West-1 during 280\,\si{\s} after a 2\,\si{\s} proton pulse of 4400\,\si{\micro \coulomb}. The Poisson error of the individual measurements and the uncertainty of the amount of sD$_2$ is smaller than the marker size. The slope (\textbf{dashed line}) and intercept (\textbf{dotted line}) from a linear fit are indicated in the plot. The intercept is consistent with the detected background rate listed in Tab.\,\ref{tab:measurement:results}.}
	\label{fig:result}
\end{figure}

A CASCADE 2D U-200 detector\footnotemark[4]
\footnotetext[3]{The volumes were taken from a fully detailed CAD model of the source and include small volumes of all transfer lines up to corresponding valves.}
\footnotetext[4]{www.n-cdt.com/cascade-2d-200}
counted UCNs for $280\,$s after receiving a trigger indicating the imminent proton beam pulse. The CASCADE detection efficiency is determined mostly by UCN losses in the 100\,\si{\micro \m} AlMg3 entrance window. The entrance window is made from the exact same raw material as the AlMg3 vacuum separation windows in the beamlines and was included in our MCUCN simulation with parameters consistent with calibration measurements (see Fig.\,28 in Ref. \cite{Bison2022}) and measured transmission through material foils \cite{Atchison2009}. The detector was mounted on a vacuum shutter at the end of a 1\,\si{\m} long Ni/Mo 85/15 \cite{Bison2022} coated glass guide, which in turn was connected to the West-1 beamport. The setup was identical to the one presented in \cite{Bison2022} used for the calibration measurements of the MCUCN simulation model of the source. The HIPA beam current during the measurements was 2.2\,\si{\milli \ampere} and the duration of the proton beam pulses was set to $2$\,\si{\s}. Thus, 4400\,\si{\micro \coulomb} protons per beam pulse were deposited on average on the spallation target, the individual values were monitored for each pulse. The UCN valve on the bottom of the storage vessel of the UCN source was permanently open. This configuration, called `norm pulse', has been used many years as a standard setting for comparison of UCN yield measurements.

\begin{table}[t]
	\centering
	\begin{tabular}{m{3.3cm} | >{\centering}m{0.8cm} >{\centering}m{0.8cm} >{\centering}m{0.8cm} | >{\centering}m{0.9cm} >{\centering\arraybackslash}m{0.9cm}}
		Simulated UCN yield & \multicolumn{3}{c|}{diffuse boost} & \multicolumn{2}{c}{perpendicular}  \\ 
		($10^{-3}\,\si{\per \cubic \centi \m} \, \si{\per \micro \coulomb}$) & A & B & C & D & E \\\hline
		& & & & & \\[-0.8em]
		from base & 36.3 & 34.6 & 29.8 & 48.3 & -\\
		from side walls & 32.5 & 32.4 & 33.3 & - & 30.3\\ \hline
		& & & & & \\[-0.9em]
		combined & 34.1 & 33.3 & 31.9 & \multicolumn{2}{c}{37.7}
	\end{tabular}
	\caption{Simulated UCN yield based on the thermal neutron flux and on different UCN transmission spectra (A - E) from the configurations with diffuse boost and boost perpendicular to the surface, as defined in section\,\ref{sec:UCNLossTransport}, Fig.\,\ref{fig:transmission}. The combined yield from the moderator vessel base and side wall is the sum weighted by the corresponding surface areas.}
	\label{tab:result}
\end{table}


\section{Analysis and discussion}
\label{sec:analysis}

The measured UCN rates at the beamport West-1 with different amounts of sD$_2$ are listed in Tab.\,\ref{tab:measurement:results}. For each pulse the proton beam current was monitored and the corresponding UCN count rate normalized to a beam current of 2.2 mA. The spread of the count rates per amount of sD$_2$ exceeds the Poisson errors in a few cases due to the influence of small fluctuations of proton beam alignment \cite{Ries2016} and other beam parameters on the UCN yield that have not been corrected for in this analysis. To estimate the count rate per mol of sD$_2$ and its uncertainty, we performed a least square fit of the individual measurements (see Fig.\,\ref{fig:result}) and obtain an intercept of $(2976 \, \pm \, 83) \, / \, 280 \,\si{\s}$, consistent with the average of the measured count rates from the empty moderator vessel. We attribute this rate to UCN and VCN (very cold neutron) production in the moderator vessel material or surrounding structure, as well as to fast neutron radiation from the spallation target during the proton beam pulse. The fitted slope of $(2923 \, \pm \, 20)\, / \, 280 \,\si{\s \per \mol}$ corresponds to a UCN yield per sD$_2$ volume and proton charge on the spallation target of
\begin{equation*}
(33.3\pm0.2) \times 10^{-3} \, \si{\per \cubic \centi \m} \, \si{\per \micro \coulomb} \quad \text{(measured)},
\end{equation*}
where we used the molar volume $V_\textsubscript{m} = 19.93 \,\si{\cubic \centi \m \per \mol}$ of sD$_2$ at approximately 5\,\si{\K} from Ref.\,\cite{Souers1986}.\\

The simulated UCN yield per sD$_2$ volume and protons on the spallation target, based on the evaluation of Eq.\,\eqref{eq:result} with the downscattering rate, Eq.\,\eqref{eq:downscatteringEtherm} and the transmission probabilities shown in Fig.\,\ref{fig:transmission}, are listed in Table\,\ref{tab:result}. We used the volume average of the thermal neutron flux (see Fig.\,\ref{fig:flux}) from the bottom to the top of the moderator vessel, which is a valid approximation with a less than 5\,\% effect on the total UCN yield when assuming a uniform deposition of the sD$_2$ on all cold moderator vessel surfaces (base and side walls). The measured UCN yield is consistent with the simulated yield within model uncertainties of approximately 10\,\% for configurations with diffuse boost and within 15\,\% for the configuration with perpendicular boost from the moderator vessel base and side wall.

\section{Conclusion}

The linearity of the measured UCN count rates at beamport West-1 as a function of the amount of sD$_2$ in the moderator vessel validates our model, Eq.\,\eqref{eq:result}. In particular, it is consistent with the assumption that the D$_2$ vapor was deposited sufficiently uniform, such that the thin film thickness is well below the mean free path of neutron losses computed from the lifetime Eq.\,\eqref{eq:lifetime} in sD$_2$. Otherwise, one would expect deviations from a linear behavior towards lower measured UCN count rates for increasing sD$_2$ filling levels. \\

We conclude that the process of spallation, thermal moderation, as well as the UCN transport to the beamports has been modelled accurately within approximately 10\,\% by our simulations (with the more likely scenario of a diffuse boost from the sD$_2$ surfaces) and previous calibration measurements \cite{Becker2015,Bison2022}. The measured UCN yield (Fig.\,\ref{fig:result}) obtained from thin films of sD$_2$ of less than 20\,\si{\mol} corroborates the calculated downscattering rate Eq.\,\eqref{eq:downscatteringEtherm} based on the thermal neutron flux (Fig.\,\ref{fig:flux}). It also provides an absolute calibration of the UCN transport from the moderator vessel to UCN detector in the experimental area. This result serves as a benchmark for future characterizations of the UCN production and extraction from sD$_2$ with filling levels in the moderator vessel of around 13\,\si{\centi \m} \cite{Bison2020} during standard operation, corresponding to approximately 1100\,\si{\mol} deuterium.

\section*{Acknowledgements}

We thank Michael Wohlmuther for programming the first version of the UCN source MCNP model. We acknowledge the PSI proton accelerator operations section, all colleagues who have been contributing to the UCN source operation at PSI, and especially the BSQ group which has been operating the PSI UCN source during the measurements, namely P. Erismann and also A. Anghel. Excellent technical support by F. Burri and M. Meier is acknowledged. This work was supported by the Swiss National Science Foundation Projects %
117696, 
137664, 
163413, 
169596, 
172626, 
178951, 
188700 
and 200441. 


\begin{thebibliography}{44}%
	\makeatletter
	\providecommand \@ifxundefined [1]{%
		\@ifx{#1\undefined}
	}%
	\providecommand \@ifnum [1]{%
		\ifnum #1\expandafter \@firstoftwo
		\else \expandafter \@secondoftwo
		\fi
	}%
	\providecommand \@ifx [1]{%
		\ifx #1\expandafter \@firstoftwo
		\else \expandafter \@secondoftwo
		\fi
	}%
	\providecommand \natexlab [1]{#1}%
	\providecommand \enquote  [1]{``#1''}%
	\providecommand \bibnamefont  [1]{#1}%
	\providecommand \bibfnamefont [1]{#1}%
	\providecommand \citenamefont [1]{#1}%
	\providecommand \href@noop [0]{\@secondoftwo}%
	\providecommand \href [0]{\begingroup \@sanitize@url \@href}%
	\providecommand \@href[1]{\@@startlink{#1}\@@href}%
	\providecommand \@@href[1]{\endgroup#1\@@endlink}%
	\providecommand \@sanitize@url [0]{\catcode `\\12\catcode `\$12\catcode
		`\&12\catcode `\#12\catcode `\^12\catcode `\_12\catcode `\%12\relax}%
	\providecommand \@@startlink[1]{}%
	\providecommand \@@endlink[0]{}%
	\providecommand \url  [0]{\begingroup\@sanitize@url \@url }%
	\providecommand \@url [1]{\endgroup\@href {#1}{\urlprefix }}%
	\providecommand \urlprefix  [0]{URL }%
	\providecommand \Eprint [0]{\href }%
	\providecommand \doibase [0]{https://doi.org/}%
	\providecommand \selectlanguage [0]{\@gobble}%
	\providecommand \bibinfo  [0]{\@secondoftwo}%
	\providecommand \bibfield  [0]{\@secondoftwo}%
	\providecommand \translation [1]{[#1]}%
	\providecommand \BibitemOpen [0]{}%
	\providecommand \bibitemStop [0]{}%
	\providecommand \bibitemNoStop [0]{.\EOS\space}%
	\providecommand \EOS [0]{\spacefactor3000\relax}%
	\providecommand \BibitemShut  [1]{\csname bibitem#1\endcsname}%
	\let\auto@bib@innerbib\@empty
	\bibitem [{\citenamefont {Zeldovich}(1959)}]{Zeldovich}%
	\BibitemOpen
	\bibfield  {author} {\bibinfo {author} {\bibfnamefont {Y.~B.}\ \bibnamefont
			{Zeldovich}},\ }\href@noop {} {\bibfield  {journal} {\bibinfo  {journal}
			{Sov. Phys. JETP-9 1389-90}\ } (\bibinfo {year} {1959})}\BibitemShut
	{NoStop}%
	\bibitem [{\citenamefont {Steyerl}(1969)}]{Steyerl1969}%
	\BibitemOpen
	\bibfield  {author} {\bibinfo {author} {\bibfnamefont {A.}~\bibnamefont
			{Steyerl}},\ }\bibfield  {title} {\bibinfo {title} {Measurement of total
			cross sections for very slow neutrons with velocities from 100m/sec to
			5m/sec},\ }\href {https://doi.org/10.1016/0370-2693(69)90127-0} {\bibfield
		{journal} {\bibinfo  {journal} {Phys. Lett.}\ }\textbf {\bibinfo {volume}
			{29b}},\ \bibinfo {pages} {33} (\bibinfo {year} {1969})}\BibitemShut
	{NoStop}%
	\bibitem [{\citenamefont {Lushchikov}\ \emph {et~al.}(1969)\citenamefont
		{Lushchikov}, \citenamefont {Pokotilovskii}, \citenamefont {Strelkov},\ and\
		\citenamefont {Shapiro}}]{Lushchikov1969}%
	\BibitemOpen
	\bibfield  {author} {\bibinfo {author} {\bibfnamefont {V.}~\bibnamefont
			{Lushchikov}}, \bibinfo {author} {\bibfnamefont {Y.}~\bibnamefont
			{Pokotilovskii}}, \bibinfo {author} {\bibfnamefont {A.}~\bibnamefont
			{Strelkov}},\ and\ \bibinfo {author} {\bibfnamefont {F.}~\bibnamefont
			{Shapiro}},\ }\bibfield  {title} {\bibinfo {title} {Observation of ultracold
			neutrons},\ }\href@noop {} {\bibfield  {journal} {\bibinfo  {journal} {JETP
				Lett.}\ }\textbf {\bibinfo {volume} {9}},\ \bibinfo {pages} {23} (\bibinfo
		{year} {1969})}\BibitemShut {NoStop}%
	\bibitem [{\citenamefont {Ignatovich}(1990)}]{Ignatovich1990}%
	\BibitemOpen
	\bibfield  {author} {\bibinfo {author} {\bibfnamefont {V.}~\bibnamefont
			{Ignatovich}},\ }\href@noop {} {\emph {\bibinfo {title} {{The Physics of
					Ultracold Neutrons}}}},\ edited by\ \bibinfo {editor} {\bibfnamefont {O.~S.}\
		\bibnamefont {Publishing}}\ (\bibinfo  {publisher} {Clarendon, Oxford},\
	\bibinfo {year} {1990})\BibitemShut {NoStop}%
	\bibitem [{\citenamefont {Golub}\ \emph {et~al.}(1991)\citenamefont {Golub},
		\citenamefont {Richardson},\ and\ \citenamefont {Lamoreaux}}]{Golub1991}%
	\BibitemOpen
	\bibfield  {author} {\bibinfo {author} {\bibfnamefont {R.}~\bibnamefont
			{Golub}}, \bibinfo {author} {\bibfnamefont {D.}~\bibnamefont {Richardson}},\
		and\ \bibinfo {author} {\bibfnamefont {S.}~\bibnamefont {Lamoreaux}},\
	}\href@noop {} {\emph {\bibinfo {title} {Ultra-Cold Neutrons}}}\ (\bibinfo
	{publisher} {Adam Hilger, Bristol, Philadelphia, and New York},\ \bibinfo
	{year} {1991})\BibitemShut {NoStop}%
	\bibitem [{\citenamefont {{Altarev}}\ \emph {et~al.}(1980)\citenamefont
		{{Altarev}}, \citenamefont {{Borisov}}, \citenamefont {{Borovikova}},
		\citenamefont {{Brandin}}, \citenamefont {{Egorov}}, \citenamefont {{Ezhov}},
		\citenamefont {{Ivanov}}, \citenamefont {{Lobashev}}, \citenamefont
		{{Nazarenko}}, \citenamefont {{Ryabov}}, \citenamefont {{Serebrov}},\ and\
		\citenamefont {{Taldaev}}}]{Altarev1980}%
	\BibitemOpen
	\bibfield  {author} {\bibinfo {author} {\bibfnamefont {I.~S.}\ \bibnamefont
			{{Altarev}}}, \bibinfo {author} {\bibfnamefont {Y.~V.}\ \bibnamefont
			{{Borisov}}}, \bibinfo {author} {\bibfnamefont {N.~V.}\ \bibnamefont
			{{Borovikova}}}, \bibinfo {author} {\bibfnamefont {A.~B.}\ \bibnamefont
			{{Brandin}}}, \bibinfo {author} {\bibfnamefont {A.~I.}\ \bibnamefont
			{{Egorov}}}, \bibinfo {author} {\bibfnamefont {V.~F.}\ \bibnamefont
			{{Ezhov}}}, \bibinfo {author} {\bibfnamefont {S.~N.}\ \bibnamefont
			{{Ivanov}}}, \bibinfo {author} {\bibfnamefont {V.~M.}\ \bibnamefont
			{{Lobashev}}}, \bibinfo {author} {\bibfnamefont {V.~A.}\ \bibnamefont
			{{Nazarenko}}}, \bibinfo {author} {\bibfnamefont {V.~L.}\ \bibnamefont
			{{Ryabov}}}, \bibinfo {author} {\bibfnamefont {A.~P.}\ \bibnamefont
			{{Serebrov}}},\ and\ \bibinfo {author} {\bibfnamefont {R.~R.}\ \bibnamefont
			{{Taldaev}}},\ }\bibfield  {title} {\bibinfo {title} {{A new upper limit on
				the electric dipole moment of the neutron}},\ }\href
	{https://doi.org/10.1016/0370-2693(81)90202-1} {\bibfield  {journal}
		{\bibinfo  {journal} {Phys. Lett. A}\ }\textbf {\bibinfo {volume} {80}},\
		\bibinfo {pages} {413} (\bibinfo {year} {1980})}\BibitemShut {NoStop}%
	\bibitem [{\citenamefont {Baker}\ \emph {et~al.}(2006)\citenamefont {Baker},
		\citenamefont {Doyle}, \citenamefont {Geltenbort}, \citenamefont {Green},
		\citenamefont {van~der Grinten}, \citenamefont {Harris}, \citenamefont
		{Iaydjiev}, \citenamefont {Ivanov}, \citenamefont {May}, \citenamefont
		{Pendlebury}, \citenamefont {Richardson}, \citenamefont {Shiers},\ and\
		\citenamefont {Smith}}]{Baker2006}%
	\BibitemOpen
	\bibfield  {author} {\bibinfo {author} {\bibfnamefont {C.~A.}\ \bibnamefont
			{Baker}}, \bibinfo {author} {\bibfnamefont {D.~D.}\ \bibnamefont {Doyle}},
		\bibinfo {author} {\bibfnamefont {P.}~\bibnamefont {Geltenbort}}, \bibinfo
		{author} {\bibfnamefont {K.}~\bibnamefont {Green}}, \bibinfo {author}
		{\bibfnamefont {M.~G.~D.}\ \bibnamefont {van~der Grinten}}, \bibinfo {author}
		{\bibfnamefont {P.~G.}\ \bibnamefont {Harris}}, \bibinfo {author}
		{\bibfnamefont {P.}~\bibnamefont {Iaydjiev}}, \bibinfo {author}
		{\bibfnamefont {S.~N.}\ \bibnamefont {Ivanov}}, \bibinfo {author}
		{\bibfnamefont {D.~J.~R.}\ \bibnamefont {May}}, \bibinfo {author}
		{\bibfnamefont {J.~M.}\ \bibnamefont {Pendlebury}}, \bibinfo {author}
		{\bibfnamefont {J.~D.}\ \bibnamefont {Richardson}}, \bibinfo {author}
		{\bibfnamefont {D.}~\bibnamefont {Shiers}},\ and\ \bibinfo {author}
		{\bibfnamefont {K.~F.}\ \bibnamefont {Smith}},\ }\bibfield  {title} {\bibinfo
		{title} {Improved experimental limit on the electric dipole moment of the
			neutron},\ }\href {https://doi.org/10.1103/PhysRevLett.97.131801} {\bibfield
		{journal} {\bibinfo  {journal} {Phys. Rev. Lett.}\ }\textbf {\bibinfo
			{volume} {97}},\ \bibinfo {pages} {131801} (\bibinfo {year}
		{2006})}\BibitemShut {NoStop}%
	\bibitem [{\citenamefont {Pendlebury}\ \emph {et~al.}(2015)\citenamefont
		{Pendlebury}, \citenamefont {Afach}, \citenamefont {Ayres}, \citenamefont
		{Baker}, \citenamefont {Ban}, \citenamefont {Bison}, \citenamefont {Bodek},
		\citenamefont {Burghoff}, \citenamefont {Geltenbort}, \citenamefont {Green},
		\citenamefont {Griffith}, \citenamefont {van~der Grinten}, \citenamefont
		{Gruji\ifmmode~\acute{c}\else \'{c}\fi{}}, \citenamefont {Harris},
		\citenamefont {H\'elaine}, \citenamefont {Iaydjiev}, \citenamefont {Ivanov},
		\citenamefont {Kasprzak}, \citenamefont {Kermaidic}, \citenamefont {Kirch},
		\citenamefont {Koch}, \citenamefont {Komposch}, \citenamefont {Kozela},
		\citenamefont {Krempel}, \citenamefont {Lauss}, \citenamefont {Lefort},
		\citenamefont {Lemi\`ere}, \citenamefont {May}, \citenamefont {Musgrave},
		\citenamefont {Naviliat-Cuncic}, \citenamefont {Piegsa}, \citenamefont
		{Pignol}, \citenamefont {Prashanth}, \citenamefont {Qu\'em\'ener},
		\citenamefont {Rawlik}, \citenamefont {Rebreyend}, \citenamefont
		{Richardson}, \citenamefont {Ries}, \citenamefont {Roccia}, \citenamefont
		{Rozpedzik}, \citenamefont {Schnabel}, \citenamefont {Schmidt-Wellenburg},
		\citenamefont {Severijns}, \citenamefont {Shiers}, \citenamefont {Thorne},
		\citenamefont {Weis}, \citenamefont {Winston}, \citenamefont {Wursten},
		\citenamefont {Zejma},\ and\ \citenamefont {Zsigmond}}]{Pendlebury2015}%
	\BibitemOpen
	\bibfield  {author} {\bibinfo {author} {\bibfnamefont {J.~M.}\ \bibnamefont
			{Pendlebury}}, \bibinfo {author} {\bibfnamefont {S.}~\bibnamefont {Afach}},
		\bibinfo {author} {\bibfnamefont {N.~J.}\ \bibnamefont {Ayres}}, \bibinfo
		{author} {\bibfnamefont {C.~A.}\ \bibnamefont {Baker}}, \bibinfo {author}
		{\bibfnamefont {G.}~\bibnamefont {Ban}}, \bibinfo {author} {\bibfnamefont
			{G.}~\bibnamefont {Bison}}, \bibinfo {author} {\bibfnamefont
			{K.}~\bibnamefont {Bodek}}, \bibinfo {author} {\bibfnamefont
			{M.}~\bibnamefont {Burghoff}}, \bibinfo {author} {\bibfnamefont
			{P.}~\bibnamefont {Geltenbort}}, \bibinfo {author} {\bibfnamefont
			{K.}~\bibnamefont {Green}}, \bibinfo {author} {\bibfnamefont {W.~C.}\
			\bibnamefont {Griffith}}, \bibinfo {author} {\bibfnamefont {M.}~\bibnamefont
			{van~der Grinten}}, \bibinfo {author} {\bibfnamefont {Z.~D.}\ \bibnamefont
			{Gruji\ifmmode~\acute{c}\else \'{c}\fi{}}}, \bibinfo {author} {\bibfnamefont
			{P.~G.}\ \bibnamefont {Harris}}, \bibinfo {author} {\bibfnamefont
			{V.}~\bibnamefont {H\'elaine}}, \bibinfo {author} {\bibfnamefont
			{P.}~\bibnamefont {Iaydjiev}}, \bibinfo {author} {\bibfnamefont {S.~N.}\
			\bibnamefont {Ivanov}}, \bibinfo {author} {\bibfnamefont {M.}~\bibnamefont
			{Kasprzak}}, \bibinfo {author} {\bibfnamefont {Y.}~\bibnamefont {Kermaidic}},
		\bibinfo {author} {\bibfnamefont {K.}~\bibnamefont {Kirch}}, \bibinfo
		{author} {\bibfnamefont {H.-C.}\ \bibnamefont {Koch}}, \bibinfo {author}
		{\bibfnamefont {S.}~\bibnamefont {Komposch}}, \bibinfo {author}
		{\bibfnamefont {A.}~\bibnamefont {Kozela}}, \bibinfo {author} {\bibfnamefont
			{J.}~\bibnamefont {Krempel}}, \bibinfo {author} {\bibfnamefont
			{B.}~\bibnamefont {Lauss}}, \bibinfo {author} {\bibfnamefont
			{T.}~\bibnamefont {Lefort}}, \bibinfo {author} {\bibfnamefont
			{Y.}~\bibnamefont {Lemi\`ere}}, \bibinfo {author} {\bibfnamefont {D.~J.~R.}\
			\bibnamefont {May}}, \bibinfo {author} {\bibfnamefont {M.}~\bibnamefont
			{Musgrave}}, \bibinfo {author} {\bibfnamefont {O.}~\bibnamefont
			{Naviliat-Cuncic}}, \bibinfo {author} {\bibfnamefont {F.~M.}\ \bibnamefont
			{Piegsa}}, \bibinfo {author} {\bibfnamefont {G.}~\bibnamefont {Pignol}},
		\bibinfo {author} {\bibfnamefont {P.~N.}\ \bibnamefont {Prashanth}}, \bibinfo
		{author} {\bibfnamefont {G.}~\bibnamefont {Qu\'em\'ener}}, \bibinfo {author}
		{\bibfnamefont {M.}~\bibnamefont {Rawlik}}, \bibinfo {author} {\bibfnamefont
			{D.}~\bibnamefont {Rebreyend}}, \bibinfo {author} {\bibfnamefont {J.~D.}\
			\bibnamefont {Richardson}}, \bibinfo {author} {\bibfnamefont
			{D.}~\bibnamefont {Ries}}, \bibinfo {author} {\bibfnamefont {S.}~\bibnamefont
			{Roccia}}, \bibinfo {author} {\bibfnamefont {D.}~\bibnamefont {Rozpedzik}},
		\bibinfo {author} {\bibfnamefont {A.}~\bibnamefont {Schnabel}}, \bibinfo
		{author} {\bibfnamefont {P.}~\bibnamefont {Schmidt-Wellenburg}}, \bibinfo
		{author} {\bibfnamefont {N.}~\bibnamefont {Severijns}}, \bibinfo {author}
		{\bibfnamefont {D.}~\bibnamefont {Shiers}}, \bibinfo {author} {\bibfnamefont
			{J.~A.}\ \bibnamefont {Thorne}}, \bibinfo {author} {\bibfnamefont
			{A.}~\bibnamefont {Weis}}, \bibinfo {author} {\bibfnamefont {O.~J.}\
			\bibnamefont {Winston}}, \bibinfo {author} {\bibfnamefont {E.}~\bibnamefont
			{Wursten}}, \bibinfo {author} {\bibfnamefont {J.}~\bibnamefont {Zejma}},\
		and\ \bibinfo {author} {\bibfnamefont {G.}~\bibnamefont {Zsigmond}},\
	}\bibfield  {title} {\bibinfo {title} {Revised experimental upper limit on
			the electric dipole moment of the neutron},\ }\href
	{https://doi.org/10.1103/PhysRevD.92.092003} {\bibfield  {journal} {\bibinfo
			{journal} {Phys. Rev. D}\ }\textbf {\bibinfo {volume} {92}},\ \bibinfo
		{pages} {092003} (\bibinfo {year} {2015})}\BibitemShut {NoStop}%
	\bibitem [{\citenamefont {Engel}\ \emph {et~al.}(2013)\citenamefont {Engel},
		\citenamefont {Ramsey-Musolf},\ and\ \citenamefont {van Kolck}}]{Engel2013a}%
	\BibitemOpen
	\bibfield  {author} {\bibinfo {author} {\bibfnamefont {J.}~\bibnamefont
			{Engel}}, \bibinfo {author} {\bibfnamefont {M.~J.}\ \bibnamefont
			{Ramsey-Musolf}},\ and\ \bibinfo {author} {\bibfnamefont {U.}~\bibnamefont
			{van Kolck}},\ }\bibfield  {title} {\bibinfo {title} {Electric dipole moments
			of nucleons, nuclei, and atoms: The standard model and beyond},\ }\href
	{https://doi.org/10.1016/j.ppnp.2013.03.003} {\bibfield  {journal} {\bibinfo
			{journal} {Progress in Particle and Nuclear Physics}\ }\textbf {\bibinfo
			{volume} {71}},\ \bibinfo {pages} {21} (\bibinfo {year} {2013})}\BibitemShut
	{NoStop}%
	\bibitem [{\citenamefont {Sakharov}(1991)}]{Sakharov1991}%
	\BibitemOpen
	\bibfield  {author} {\bibinfo {author} {\bibfnamefont {A.~D.}\ \bibnamefont
			{Sakharov}},\ }\bibfield  {title} {\bibinfo {title} {Violation of cp
			invariance, c asymmetry, and baryon asymmetry of the universe},\ }\href
	{https://doi.org/10.1070/pu1991v034n05abeh002497} {\bibfield  {journal}
		{\bibinfo  {journal} {Soviet Physics Uspekhi}\ }\textbf {\bibinfo {volume}
			{34}},\ \bibinfo {pages} {392} (\bibinfo {year} {1991})}\BibitemShut
	{NoStop}%
	\bibitem [{\citenamefont {Morrissey}\ and\ \citenamefont
		{Ramsey-Musolf}(2012)}]{Morrissey2012}%
	\BibitemOpen
	\bibfield  {author} {\bibinfo {author} {\bibfnamefont {D.~E.}\ \bibnamefont
			{Morrissey}}\ and\ \bibinfo {author} {\bibfnamefont {M.~J.}\ \bibnamefont
			{Ramsey-Musolf}},\ }\bibfield  {title} {\bibinfo {title} {Electroweak
			baryogenesis},\ }\href {https://doi.org/10.1088/1367-2630/14/12/125003}
	{\bibfield  {journal} {\bibinfo  {journal} {New Journal of Physics}\ }\textbf
		{\bibinfo {volume} {14}},\ \bibinfo {pages} {125003} (\bibinfo {year}
		{2012})}\BibitemShut {NoStop}%
	\bibitem [{\citenamefont {Abel}\ \emph {et~al.}(2020)\citenamefont {Abel},
		\citenamefont {Afach}, \citenamefont {Ayres}, \citenamefont {Baker},
		\citenamefont {Ban}, \citenamefont {Bison}, \citenamefont {Bodek},
		\citenamefont {Bondar}, \citenamefont {Burghoff}, \citenamefont {Chanel},
		\citenamefont {Chowdhuri}, \citenamefont {Chiu}, \citenamefont {Clement},
		\citenamefont {Crawford}, \citenamefont {Daum}, \citenamefont {Emmenegger},
		\citenamefont {Ferraris-Bouchez}, \citenamefont {Fertl}, \citenamefont
		{Flaux}, \citenamefont {Franke}, \citenamefont {Fratangelo}, \citenamefont
		{Geltenbort}, \citenamefont {Green}, \citenamefont {Griffith}, \citenamefont
		{van{\hspace{0.167em}}der{\hspace{0.167em}}Grinten}, \citenamefont
		{Gruji{\'{c}}}, \citenamefont {Harris}, \citenamefont {Hayen}, \citenamefont
		{Heil}, \citenamefont {Henneck}, \citenamefont {H{\'{e}}laine}, \citenamefont
		{Hild}, \citenamefont {Hodge}, \citenamefont {Horras}, \citenamefont
		{Iaydjiev}, \citenamefont {Ivanov}, \citenamefont {Kasprzak}, \citenamefont
		{Kermaidic}, \citenamefont {Kirch}, \citenamefont {Knecht}, \citenamefont
		{Knowles}, \citenamefont {Koch}, \citenamefont {Koss}, \citenamefont
		{Komposch}, \citenamefont {Kozela}, \citenamefont {Kraft}, \citenamefont
		{Krempel}, \citenamefont {Ku{\'{z}}niak}, \citenamefont {Lauss},
		\citenamefont {Lefort}, \citenamefont {Lemi{\`{e}}re}, \citenamefont
		{Leredde}, \citenamefont {Mohanmurthy}, \citenamefont {Mtchedlishvili},
		\citenamefont {Musgrave}, \citenamefont {Naviliat-Cuncic}, \citenamefont
		{Pais}, \citenamefont {Piegsa}, \citenamefont {Pierre}, \citenamefont
		{Pignol}, \citenamefont {Plonka-Spehr}, \citenamefont {Prashanth},
		\citenamefont {Qu{\'{e}}m{\'{e}}ner}, \citenamefont {Rawlik}, \citenamefont
		{Rebreyend}, \citenamefont {Rien{\"{a}}cker}, \citenamefont {Ries},
		\citenamefont {Roccia}, \citenamefont {Rogel}, \citenamefont {Rozpedzik},
		\citenamefont {Schnabel}, \citenamefont {Schmidt-Wellenburg}, \citenamefont
		{Severijns}, \citenamefont {Shiers}, \citenamefont {Dinani}, \citenamefont
		{Thorne}, \citenamefont {Virot}, \citenamefont {Voigt}, \citenamefont {Weis},
		\citenamefont {Wursten}, \citenamefont {Wyszynski}, \citenamefont {Zejma},
		\citenamefont {Zenner},\ and\ \citenamefont {Zsigmond}}]{Abel2020}%
	\BibitemOpen
	\bibfield  {author} {\bibinfo {author} {\bibfnamefont {C.}~\bibnamefont
			{Abel}}, \bibinfo {author} {\bibfnamefont {S.}~\bibnamefont {Afach}},
		\bibinfo {author} {\bibfnamefont {N.}~\bibnamefont {Ayres}}, \bibinfo
		{author} {\bibfnamefont {C.}~\bibnamefont {Baker}}, \bibinfo {author}
		{\bibfnamefont {G.}~\bibnamefont {Ban}}, \bibinfo {author} {\bibfnamefont
			{G.}~\bibnamefont {Bison}}, \bibinfo {author} {\bibfnamefont
			{K.}~\bibnamefont {Bodek}}, \bibinfo {author} {\bibfnamefont
			{V.}~\bibnamefont {Bondar}}, \bibinfo {author} {\bibfnamefont
			{M.}~\bibnamefont {Burghoff}}, \bibinfo {author} {\bibfnamefont
			{E.}~\bibnamefont {Chanel}}, \bibinfo {author} {\bibfnamefont
			{Z.}~\bibnamefont {Chowdhuri}}, \bibinfo {author} {\bibfnamefont {P.-J.}\
			\bibnamefont {Chiu}}, \bibinfo {author} {\bibfnamefont {B.}~\bibnamefont
			{Clement}}, \bibinfo {author} {\bibfnamefont {C.}~\bibnamefont {Crawford}},
		\bibinfo {author} {\bibfnamefont {M.}~\bibnamefont {Daum}}, \bibinfo {author}
		{\bibfnamefont {S.}~\bibnamefont {Emmenegger}}, \bibinfo {author}
		{\bibfnamefont {L.}~\bibnamefont {Ferraris-Bouchez}}, \bibinfo {author}
		{\bibfnamefont {M.}~\bibnamefont {Fertl}}, \bibinfo {author} {\bibfnamefont
			{P.}~\bibnamefont {Flaux}}, \bibinfo {author} {\bibfnamefont
			{B.}~\bibnamefont {Franke}}, \bibinfo {author} {\bibfnamefont
			{A.}~\bibnamefont {Fratangelo}}, \bibinfo {author} {\bibfnamefont
			{P.}~\bibnamefont {Geltenbort}}, \bibinfo {author} {\bibfnamefont
			{K.}~\bibnamefont {Green}}, \bibinfo {author} {\bibfnamefont
			{W.}~\bibnamefont {Griffith}}, \bibinfo {author} {\bibfnamefont
			{M.}~\bibnamefont {van{\hspace{0.167em}}der{\hspace{0.167em}}Grinten}},
		\bibinfo {author} {\bibfnamefont {Z.}~\bibnamefont {Gruji{\'{c}}}}, \bibinfo
		{author} {\bibfnamefont {P.}~\bibnamefont {Harris}}, \bibinfo {author}
		{\bibfnamefont {L.}~\bibnamefont {Hayen}}, \bibinfo {author} {\bibfnamefont
			{W.}~\bibnamefont {Heil}}, \bibinfo {author} {\bibfnamefont {R.}~\bibnamefont
			{Henneck}}, \bibinfo {author} {\bibfnamefont {V.}~\bibnamefont
			{H{\'{e}}laine}}, \bibinfo {author} {\bibfnamefont {N.}~\bibnamefont {Hild}},
		\bibinfo {author} {\bibfnamefont {Z.}~\bibnamefont {Hodge}}, \bibinfo
		{author} {\bibfnamefont {M.}~\bibnamefont {Horras}}, \bibinfo {author}
		{\bibfnamefont {P.}~\bibnamefont {Iaydjiev}}, \bibinfo {author}
		{\bibfnamefont {S.}~\bibnamefont {Ivanov}}, \bibinfo {author} {\bibfnamefont
			{M.}~\bibnamefont {Kasprzak}}, \bibinfo {author} {\bibfnamefont
			{Y.}~\bibnamefont {Kermaidic}}, \bibinfo {author} {\bibfnamefont
			{K.}~\bibnamefont {Kirch}}, \bibinfo {author} {\bibfnamefont
			{A.}~\bibnamefont {Knecht}}, \bibinfo {author} {\bibfnamefont
			{P.}~\bibnamefont {Knowles}}, \bibinfo {author} {\bibfnamefont {H.-C.}\
			\bibnamefont {Koch}}, \bibinfo {author} {\bibfnamefont {P.}~\bibnamefont
			{Koss}}, \bibinfo {author} {\bibfnamefont {S.}~\bibnamefont {Komposch}},
		\bibinfo {author} {\bibfnamefont {A.}~\bibnamefont {Kozela}}, \bibinfo
		{author} {\bibfnamefont {A.}~\bibnamefont {Kraft}}, \bibinfo {author}
		{\bibfnamefont {J.}~\bibnamefont {Krempel}}, \bibinfo {author} {\bibfnamefont
			{M.}~\bibnamefont {Ku{\'{z}}niak}}, \bibinfo {author} {\bibfnamefont
			{B.}~\bibnamefont {Lauss}}, \bibinfo {author} {\bibfnamefont
			{T.}~\bibnamefont {Lefort}}, \bibinfo {author} {\bibfnamefont
			{Y.}~\bibnamefont {Lemi{\`{e}}re}}, \bibinfo {author} {\bibfnamefont
			{A.}~\bibnamefont {Leredde}}, \bibinfo {author} {\bibfnamefont
			{P.}~\bibnamefont {Mohanmurthy}}, \bibinfo {author} {\bibfnamefont
			{A.}~\bibnamefont {Mtchedlishvili}}, \bibinfo {author} {\bibfnamefont
			{M.}~\bibnamefont {Musgrave}}, \bibinfo {author} {\bibfnamefont
			{O.}~\bibnamefont {Naviliat-Cuncic}}, \bibinfo {author} {\bibfnamefont
			{D.}~\bibnamefont {Pais}}, \bibinfo {author} {\bibfnamefont {F.}~\bibnamefont
			{Piegsa}}, \bibinfo {author} {\bibfnamefont {E.}~\bibnamefont {Pierre}},
		\bibinfo {author} {\bibfnamefont {G.}~\bibnamefont {Pignol}}, \bibinfo
		{author} {\bibfnamefont {C.}~\bibnamefont {Plonka-Spehr}}, \bibinfo {author}
		{\bibfnamefont {P.}~\bibnamefont {Prashanth}}, \bibinfo {author}
		{\bibfnamefont {G.}~\bibnamefont {Qu{\'{e}}m{\'{e}}ner}}, \bibinfo {author}
		{\bibfnamefont {M.}~\bibnamefont {Rawlik}}, \bibinfo {author} {\bibfnamefont
			{D.}~\bibnamefont {Rebreyend}}, \bibinfo {author} {\bibfnamefont
			{I.}~\bibnamefont {Rien{\"{a}}cker}}, \bibinfo {author} {\bibfnamefont
			{D.}~\bibnamefont {Ries}}, \bibinfo {author} {\bibfnamefont {S.}~\bibnamefont
			{Roccia}}, \bibinfo {author} {\bibfnamefont {G.}~\bibnamefont {Rogel}},
		\bibinfo {author} {\bibfnamefont {D.}~\bibnamefont {Rozpedzik}}, \bibinfo
		{author} {\bibfnamefont {A.}~\bibnamefont {Schnabel}}, \bibinfo {author}
		{\bibfnamefont {P.}~\bibnamefont {Schmidt-Wellenburg}}, \bibinfo {author}
		{\bibfnamefont {N.}~\bibnamefont {Severijns}}, \bibinfo {author}
		{\bibfnamefont {D.}~\bibnamefont {Shiers}}, \bibinfo {author} {\bibfnamefont
			{R.~T.}\ \bibnamefont {Dinani}}, \bibinfo {author} {\bibfnamefont
			{J.}~\bibnamefont {Thorne}}, \bibinfo {author} {\bibfnamefont
			{R.}~\bibnamefont {Virot}}, \bibinfo {author} {\bibfnamefont
			{J.}~\bibnamefont {Voigt}}, \bibinfo {author} {\bibfnamefont
			{A.}~\bibnamefont {Weis}}, \bibinfo {author} {\bibfnamefont {E.}~\bibnamefont
			{Wursten}}, \bibinfo {author} {\bibfnamefont {G.}~\bibnamefont {Wyszynski}},
		\bibinfo {author} {\bibfnamefont {J.}~\bibnamefont {Zejma}}, \bibinfo
		{author} {\bibfnamefont {J.}~\bibnamefont {Zenner}},\ and\ \bibinfo {author}
		{\bibfnamefont {G.}~\bibnamefont {Zsigmond}},\ }\bibfield  {title} {\bibinfo
		{title} {Measurement of the permanent electric dipole moment of the
			neutron},\ }\href {https://doi.org/10.1103/physrevlett.124.081803} {\bibfield
		{journal} {\bibinfo  {journal} {Physical Review Letters}\ }\textbf {\bibinfo
			{volume} {124}},\ \bibinfo {pages} {081803} (\bibinfo {year}
		{2020})}\BibitemShut {NoStop}%
	\bibitem [{\citenamefont {Bison}\ \emph {et~al.}(2020)\citenamefont {Bison},
		\citenamefont {Blau}, \citenamefont {Daum}, \citenamefont {Göltl},
		\citenamefont {Henneck}, \citenamefont {Kirch}, \citenamefont {Lauss},
		\citenamefont {Ries}, \citenamefont {Schmidt-Wellenburg},\ and\ \citenamefont
		{Zsigmond}}]{Bison2020}%
	\BibitemOpen
	\bibfield  {author} {\bibinfo {author} {\bibfnamefont {G.}~\bibnamefont
			{Bison}}, \bibinfo {author} {\bibfnamefont {B.}~\bibnamefont {Blau}},
		\bibinfo {author} {\bibfnamefont {M.}~\bibnamefont {Daum}}, \bibinfo {author}
		{\bibfnamefont {L.}~\bibnamefont {Göltl}}, \bibinfo {author} {\bibfnamefont
			{R.}~\bibnamefont {Henneck}}, \bibinfo {author} {\bibfnamefont
			{K.}~\bibnamefont {Kirch}}, \bibinfo {author} {\bibfnamefont
			{B.}~\bibnamefont {Lauss}}, \bibinfo {author} {\bibfnamefont
			{D.}~\bibnamefont {Ries}}, \bibinfo {author} {\bibfnamefont {P.}~\bibnamefont
			{Schmidt-Wellenburg}},\ and\ \bibinfo {author} {\bibfnamefont
			{G.}~\bibnamefont {Zsigmond}},\ }\bibfield  {title} {\bibinfo {title}
		{Neutron optics of the {PSI} ultracold-neutron source: characterization and
			simulation},\ }\bibfield  {journal} {\bibinfo  {journal} {The European
			Physical Journal A}\ }\textbf {\bibinfo {volume} {56}},\ \href
	{https://doi.org/10.1140/epja/s10050-020-00027-w}
	{10.1140/epja/s10050-020-00027-w} (\bibinfo {year} {2020})\BibitemShut
	{NoStop}%
	\bibitem [{\citenamefont {Ayres}\ \emph {et~al.}(2021)\citenamefont {Ayres},
		\citenamefont {Ban}, \citenamefont {Bienstman}, \citenamefont {Bison},
		\citenamefont {Bodek}, \citenamefont {Bondar}, \citenamefont {Bouillaud},
		\citenamefont {Chanel}, \citenamefont {Chen}, \citenamefont {Chiu},
		\citenamefont {Cl{\'{e}}ment}, \citenamefont {Crawford}, \citenamefont
		{Daum}, \citenamefont {Dechenaux}, \citenamefont {Doorenbos}, \citenamefont
		{Emmenegger}, \citenamefont {Ferraris-Bouchez}, \citenamefont {Fertl},
		\citenamefont {Fratangelo}, \citenamefont {Flaux}, \citenamefont
		{Goupilli{\`{e}}re}, \citenamefont {Griffith}, \citenamefont {Grujic},
		\citenamefont {Harris}, \citenamefont {Kirch}, \citenamefont {Koss},
		\citenamefont {Krempel}, \citenamefont {Lauss}, \citenamefont {Lefort},
		\citenamefont {Lemi{\`{e}}re}, \citenamefont {Leredde}, \citenamefont
		{Meier}, \citenamefont {Menu}, \citenamefont {Mullins}, \citenamefont
		{Naviliat-Cuncic}, \citenamefont {Pais}, \citenamefont {Piegsa},
		\citenamefont {Pignol}, \citenamefont {Qu{\'{e}}m{\'{e}}ner}, \citenamefont
		{Rawlik}, \citenamefont {Rebreyend}, \citenamefont {Rienäcker}, \citenamefont
		{Ries}, \citenamefont {Roccia}, \citenamefont {Ross}, \citenamefont
		{Rozpedzik}, \citenamefont {Saenz}, \citenamefont {Schmidt-Wellenburg},
		\citenamefont {Schnabel}, \citenamefont {Severijns}, \citenamefont {Shen},
		\citenamefont {Stapf}, \citenamefont {Svirina}, \citenamefont {Dinani},
		\citenamefont {Touati}, \citenamefont {Thorne}, \citenamefont {Virot},
		\citenamefont {Voigt}, \citenamefont {Wursten}, \citenamefont {Yazdandoost},
		\citenamefont {Zejma},\ and\ \citenamefont {Zsigmond}}]{Ayres2021n2edm}%
	\BibitemOpen
	\bibfield  {author} {\bibinfo {author} {\bibfnamefont {N.~J.}\ \bibnamefont
			{Ayres}}, \bibinfo {author} {\bibfnamefont {G.}~\bibnamefont {Ban}}, \bibinfo
		{author} {\bibfnamefont {L.}~\bibnamefont {Bienstman}}, \bibinfo {author}
		{\bibfnamefont {G.}~\bibnamefont {Bison}}, \bibinfo {author} {\bibfnamefont
			{K.}~\bibnamefont {Bodek}}, \bibinfo {author} {\bibfnamefont
			{V.}~\bibnamefont {Bondar}}, \bibinfo {author} {\bibfnamefont
			{T.}~\bibnamefont {Bouillaud}}, \bibinfo {author} {\bibfnamefont
			{E.}~\bibnamefont {Chanel}}, \bibinfo {author} {\bibfnamefont
			{J.}~\bibnamefont {Chen}}, \bibinfo {author} {\bibfnamefont {P.-J.}\
			\bibnamefont {Chiu}}, \bibinfo {author} {\bibfnamefont {B.}~\bibnamefont
			{Cl{\'{e}}ment}}, \bibinfo {author} {\bibfnamefont {C.~B.}\ \bibnamefont
			{Crawford}}, \bibinfo {author} {\bibfnamefont {M.}~\bibnamefont {Daum}},
		\bibinfo {author} {\bibfnamefont {B.}~\bibnamefont {Dechenaux}}, \bibinfo
		{author} {\bibfnamefont {C.~B.}\ \bibnamefont {Doorenbos}}, \bibinfo {author}
		{\bibfnamefont {S.}~\bibnamefont {Emmenegger}}, \bibinfo {author}
		{\bibfnamefont {L.}~\bibnamefont {Ferraris-Bouchez}}, \bibinfo {author}
		{\bibfnamefont {M.}~\bibnamefont {Fertl}}, \bibinfo {author} {\bibfnamefont
			{A.}~\bibnamefont {Fratangelo}}, \bibinfo {author} {\bibfnamefont
			{P.}~\bibnamefont {Flaux}}, \bibinfo {author} {\bibfnamefont
			{D.}~\bibnamefont {Goupilli{\`{e}}re}}, \bibinfo {author} {\bibfnamefont
			{W.~C.}\ \bibnamefont {Griffith}}, \bibinfo {author} {\bibfnamefont {Z.~D.}\
			\bibnamefont {Grujic}}, \bibinfo {author} {\bibfnamefont {P.~G.}\
			\bibnamefont {Harris}}, \bibinfo {author} {\bibfnamefont {K.}~\bibnamefont
			{Kirch}}, \bibinfo {author} {\bibfnamefont {P.~A.}\ \bibnamefont {Koss}},
		\bibinfo {author} {\bibfnamefont {J.}~\bibnamefont {Krempel}}, \bibinfo
		{author} {\bibfnamefont {B.}~\bibnamefont {Lauss}}, \bibinfo {author}
		{\bibfnamefont {T.}~\bibnamefont {Lefort}}, \bibinfo {author} {\bibfnamefont
			{Y.}~\bibnamefont {Lemi{\`{e}}re}}, \bibinfo {author} {\bibfnamefont
			{A.}~\bibnamefont {Leredde}}, \bibinfo {author} {\bibfnamefont
			{M.}~\bibnamefont {Meier}}, \bibinfo {author} {\bibfnamefont
			{J.}~\bibnamefont {Menu}}, \bibinfo {author} {\bibfnamefont {D.~A.}\
			\bibnamefont {Mullins}}, \bibinfo {author} {\bibfnamefont {O.}~\bibnamefont
			{Naviliat-Cuncic}}, \bibinfo {author} {\bibfnamefont {D.}~\bibnamefont
			{Pais}}, \bibinfo {author} {\bibfnamefont {F.~M.}\ \bibnamefont {Piegsa}},
		\bibinfo {author} {\bibfnamefont {G.}~\bibnamefont {Pignol}}, \bibinfo
		{author} {\bibfnamefont {G.}~\bibnamefont {Qu{\'{e}}m{\'{e}}ner}}, \bibinfo
		{author} {\bibfnamefont {M.}~\bibnamefont {Rawlik}}, \bibinfo {author}
		{\bibfnamefont {D.}~\bibnamefont {Rebreyend}}, \bibinfo {author}
		{\bibfnamefont {I.}~\bibnamefont {Rienäcker}}, \bibinfo {author}
		{\bibfnamefont {D.}~\bibnamefont {Ries}}, \bibinfo {author} {\bibfnamefont
			{S.}~\bibnamefont {Roccia}}, \bibinfo {author} {\bibfnamefont {K.~U.}\
			\bibnamefont {Ross}}, \bibinfo {author} {\bibfnamefont {D.}~\bibnamefont
			{Rozpedzik}}, \bibinfo {author} {\bibfnamefont {W.}~\bibnamefont {Saenz}},
		\bibinfo {author} {\bibfnamefont {P.}~\bibnamefont {Schmidt-Wellenburg}},
		\bibinfo {author} {\bibfnamefont {A.}~\bibnamefont {Schnabel}}, \bibinfo
		{author} {\bibfnamefont {N.}~\bibnamefont {Severijns}}, \bibinfo {author}
		{\bibfnamefont {B.}~\bibnamefont {Shen}}, \bibinfo {author} {\bibfnamefont
			{T.}~\bibnamefont {Stapf}}, \bibinfo {author} {\bibfnamefont
			{K.}~\bibnamefont {Svirina}}, \bibinfo {author} {\bibfnamefont {R.~T.}\
			\bibnamefont {Dinani}}, \bibinfo {author} {\bibfnamefont {S.}~\bibnamefont
			{Touati}}, \bibinfo {author} {\bibfnamefont {J.}~\bibnamefont {Thorne}},
		\bibinfo {author} {\bibfnamefont {R.}~\bibnamefont {Virot}}, \bibinfo
		{author} {\bibfnamefont {J.}~\bibnamefont {Voigt}}, \bibinfo {author}
		{\bibfnamefont {E.}~\bibnamefont {Wursten}}, \bibinfo {author} {\bibfnamefont
			{N.}~\bibnamefont {Yazdandoost}}, \bibinfo {author} {\bibfnamefont
			{J.}~\bibnamefont {Zejma}},\ and\ \bibinfo {author} {\bibfnamefont
			{G.}~\bibnamefont {Zsigmond}},\ }\bibfield  {title} {\bibinfo {title} {The
			design of the n2edm experiment},\ }\bibfield  {journal} {\bibinfo  {journal}
		{The European Physical Journal C}\ }\textbf {\bibinfo {volume} {81}},\ \href
	{https://doi.org/10.1140/epjc/s10052-021-09298-z}
	{10.1140/epjc/s10052-021-09298-z} (\bibinfo {year} {2021})\BibitemShut
	{NoStop}%
	\bibitem [{\citenamefont {Wohlmuther}\ and\ \citenamefont
		{Heidenreich}(2006)}]{Wohlmuther2006}%
	\BibitemOpen
	\bibfield  {author} {\bibinfo {author} {\bibfnamefont {M.}~\bibnamefont
			{Wohlmuther}}\ and\ \bibinfo {author} {\bibfnamefont {G.}~\bibnamefont
			{Heidenreich}},\ }\bibfield  {title} {\bibinfo {title} {{The spallation
				target of the ultra-cold neutron source UCN at PSI}},\ }\href
	{https://doi.org/10.1016/j.nima.2006.03.040} {\bibfield  {journal} {\bibinfo
			{journal} {Nucl. Instrum. Methods A}\ }\textbf {\bibinfo {volume} {564}},\
		\bibinfo {pages} {51} (\bibinfo {year} {2006})}\BibitemShut {NoStop}%
	\bibitem [{\citenamefont {Grillenberger}\ \emph {et~al.}(2021)\citenamefont
		{Grillenberger}, \citenamefont {Baumgarten},\ and\ \citenamefont
		{Seidel}}]{Grillenberger2021}%
	\BibitemOpen
	\bibfield  {author} {\bibinfo {author} {\bibfnamefont {J.}~\bibnamefont
			{Grillenberger}}, \bibinfo {author} {\bibfnamefont {C.}~\bibnamefont
			{Baumgarten}},\ and\ \bibinfo {author} {\bibfnamefont {M.}~\bibnamefont
			{Seidel}},\ }\bibfield  {title} {\bibinfo {title} {The high intensity proton
			accelerator facility},\ }\bibfield  {journal} {\bibinfo  {journal} {{SciPost}
			Physics Proceedings}\ }\href {https://doi.org/10.21468/scipostphysproc.5.002}
	{10.21468/scipostphysproc.5.002} (\bibinfo {year} {2021})\BibitemShut
	{NoStop}%
	\bibitem [{\citenamefont {Golub}\ and\ \citenamefont
		{Boening}(1983)}]{Golub1983}%
	\BibitemOpen
	\bibfield  {author} {\bibinfo {author} {\bibfnamefont {R.}~\bibnamefont
			{Golub}}\ and\ \bibinfo {author} {\bibfnamefont {K.}~\bibnamefont
			{Boening}},\ }\bibfield  {title} {\bibinfo {title} {New type of low
			temperature source of ultra-cold neutrons and production of continous beams
			of {UCN}},\ }\href {https://doi.org/10.1007/bf01308763} {\bibfield  {journal}
		{\bibinfo  {journal} {Z. Phys. B}\ }\textbf {\bibinfo {volume} {51}},\
		\bibinfo {pages} {95} (\bibinfo {year} {1983})}\BibitemShut {NoStop}%
	\bibitem [{\citenamefont {Bison}\ \emph {et~al.}(2022)\citenamefont {Bison},
		\citenamefont {Daum}, \citenamefont {Kirch}, \citenamefont {Lauss},
		\citenamefont {Ries}, \citenamefont {Schmidt-Wellenburg},\ and\ \citenamefont
		{Zsigmond}}]{Bison2022}%
	\BibitemOpen
	\bibfield  {author} {\bibinfo {author} {\bibfnamefont {G.}~\bibnamefont
			{Bison}}, \bibinfo {author} {\bibfnamefont {M.}~\bibnamefont {Daum}},
		\bibinfo {author} {\bibfnamefont {K.}~\bibnamefont {Kirch}}, \bibinfo
		{author} {\bibfnamefont {B.}~\bibnamefont {Lauss}}, \bibinfo {author}
		{\bibfnamefont {D.}~\bibnamefont {Ries}}, \bibinfo {author} {\bibfnamefont
			{P.}~\bibnamefont {Schmidt-Wellenburg}},\ and\ \bibinfo {author}
		{\bibfnamefont {G.}~\bibnamefont {Zsigmond}},\ }\bibfield  {title} {\bibinfo
		{title} {Ultracold neutron storage and transport at the {PSI} {UCN} source},\
	}\bibfield  {journal} {\bibinfo  {journal} {The European Physical Journal A}\
	}\textbf {\bibinfo {volume} {58}},\ \href
	{https://doi.org/10.1140/epja/s10050-022-00747-1}
	{10.1140/epja/s10050-022-00747-1} (\bibinfo {year} {2022})\BibitemShut
	{NoStop}%
	\bibitem [{\citenamefont {Liu}\ \emph {et~al.}(2000)\citenamefont {Liu},
		\citenamefont {Young},\ and\ \citenamefont {Lamoreaux}}]{Liu2000}%
	\BibitemOpen
	\bibfield  {author} {\bibinfo {author} {\bibfnamefont {C.-Y.}\ \bibnamefont
			{Liu}}, \bibinfo {author} {\bibfnamefont {A.~R.}\ \bibnamefont {Young}},\
		and\ \bibinfo {author} {\bibfnamefont {S.~K.}\ \bibnamefont {Lamoreaux}},\
	}\bibfield  {title} {\bibinfo {title} {Ultracold neutron upscattering rates
			in a molecular deuterium crystal},\ }\href
	{https://doi.org/10.1103/PhysRevB.62.R3581} {\bibfield  {journal} {\bibinfo
			{journal} {Phys. Rev. B}\ }\textbf {\bibinfo {volume} {62}},\ \bibinfo
		{pages} {R3581} (\bibinfo {year} {2000})}\BibitemShut {NoStop}%
	\bibitem [{\citenamefont {Morris}\ \emph {et~al.}(2002)\citenamefont {Morris},
		\citenamefont {Anaya}, \citenamefont {Bowles}, \citenamefont {Filippone},
		\citenamefont {Geltenbort}, \citenamefont {Hill}, \citenamefont {Hino},
		\citenamefont {Hoedl}, \citenamefont {Hogan}, \citenamefont {Ito},
		\citenamefont {Kawai}, \citenamefont {Kirch}, \citenamefont {Lamoreaux},
		\citenamefont {Liu}, \citenamefont {Makela}, \citenamefont {Marek},
		\citenamefont {Martin}, \citenamefont {Mortensen}, \citenamefont
		{Pichlmaier}, \citenamefont {Saunders}, \citenamefont {Seestrom},
		\citenamefont {Smith}, \citenamefont {Teasdale}, \citenamefont {Tipton},
		\citenamefont {Utsuro}, \citenamefont {Young},\ and\ \citenamefont
		{Yuan}}]{Morris2002}%
	\BibitemOpen
	\bibfield  {author} {\bibinfo {author} {\bibfnamefont {C.~L.}\ \bibnamefont
			{Morris}}, \bibinfo {author} {\bibfnamefont {J.~M.}\ \bibnamefont {Anaya}},
		\bibinfo {author} {\bibfnamefont {T.~J.}\ \bibnamefont {Bowles}}, \bibinfo
		{author} {\bibfnamefont {B.~W.}\ \bibnamefont {Filippone}}, \bibinfo {author}
		{\bibfnamefont {P.}~\bibnamefont {Geltenbort}}, \bibinfo {author}
		{\bibfnamefont {R.~E.}\ \bibnamefont {Hill}}, \bibinfo {author}
		{\bibfnamefont {M.}~\bibnamefont {Hino}}, \bibinfo {author} {\bibfnamefont
			{S.}~\bibnamefont {Hoedl}}, \bibinfo {author} {\bibfnamefont {G.~E.}\
			\bibnamefont {Hogan}}, \bibinfo {author} {\bibfnamefont {T.~M.}\ \bibnamefont
			{Ito}}, \bibinfo {author} {\bibfnamefont {T.}~\bibnamefont {Kawai}}, \bibinfo
		{author} {\bibfnamefont {K.}~\bibnamefont {Kirch}}, \bibinfo {author}
		{\bibfnamefont {S.~K.}\ \bibnamefont {Lamoreaux}}, \bibinfo {author}
		{\bibfnamefont {C.-Y.}\ \bibnamefont {Liu}}, \bibinfo {author} {\bibfnamefont
			{M.}~\bibnamefont {Makela}}, \bibinfo {author} {\bibfnamefont {L.~J.}\
			\bibnamefont {Marek}}, \bibinfo {author} {\bibfnamefont {J.~W.}\ \bibnamefont
			{Martin}}, \bibinfo {author} {\bibfnamefont {R.~N.}\ \bibnamefont
			{Mortensen}}, \bibinfo {author} {\bibfnamefont {A.}~\bibnamefont
			{Pichlmaier}}, \bibinfo {author} {\bibfnamefont {A.}~\bibnamefont
			{Saunders}}, \bibinfo {author} {\bibfnamefont {S.~J.}\ \bibnamefont
			{Seestrom}}, \bibinfo {author} {\bibfnamefont {D.}~\bibnamefont {Smith}},
		\bibinfo {author} {\bibfnamefont {W.}~\bibnamefont {Teasdale}}, \bibinfo
		{author} {\bibfnamefont {B.}~\bibnamefont {Tipton}}, \bibinfo {author}
		{\bibfnamefont {M.}~\bibnamefont {Utsuro}}, \bibinfo {author} {\bibfnamefont
			{A.~R.}\ \bibnamefont {Young}},\ and\ \bibinfo {author} {\bibfnamefont
			{J.}~\bibnamefont {Yuan}},\ }\bibfield  {title} {\bibinfo {title}
		{Measurements of ultracold-neutron lifetimes in solid deuterium},\ }\href
	{https://doi.org/10.1103/PhysRevLett.89.272501} {\bibfield  {journal}
		{\bibinfo  {journal} {Phys. Rev. Lett.}\ }\textbf {\bibinfo {volume} {89}},\
		\bibinfo {pages} {272501} (\bibinfo {year} {2002})}\BibitemShut {NoStop}%
	\bibitem [{\citenamefont {Atchison}\ \emph {et~al.}(2005)\citenamefont
		{Atchison}, \citenamefont {Blau}, \citenamefont {van~den Brandt},
		\citenamefont {Brys}, \citenamefont {Daum}, \citenamefont {Fierlinger},
		\citenamefont {Hautle}, \citenamefont {Henneck}, \citenamefont {Heule},
		\citenamefont {Kasprzak}, \citenamefont {Kirch}, \citenamefont {Kohlbrecher},
		\citenamefont {Kuehen}, \citenamefont {Konter}, \citenamefont {Pichlmaier},
		\citenamefont {Wokaun}, \citenamefont {Bodek}, \citenamefont {Geltenbort},\
		and\ \citenamefont {Zmeskal}}]{Atchison2005b}%
	\BibitemOpen
	\bibfield  {author} {\bibinfo {author} {\bibfnamefont {F.}~\bibnamefont
			{Atchison}}, \bibinfo {author} {\bibfnamefont {B.}~\bibnamefont {Blau}},
		\bibinfo {author} {\bibfnamefont {B.}~\bibnamefont {van~den Brandt}},
		\bibinfo {author} {\bibfnamefont {T.}~\bibnamefont {Brys}}, \bibinfo {author}
		{\bibfnamefont {M.}~\bibnamefont {Daum}}, \bibinfo {author} {\bibfnamefont
			{P.}~\bibnamefont {Fierlinger}}, \bibinfo {author} {\bibfnamefont
			{P.}~\bibnamefont {Hautle}}, \bibinfo {author} {\bibfnamefont
			{R.}~\bibnamefont {Henneck}}, \bibinfo {author} {\bibfnamefont
			{S.}~\bibnamefont {Heule}}, \bibinfo {author} {\bibfnamefont
			{K.}~\bibnamefont {Kirch}}, \bibinfo {author} {\bibfnamefont
			{J.}~\bibnamefont {Kohlbrecher}}, \bibinfo {author} {\bibfnamefont
			{G.}~\bibnamefont {Kühne}}, \bibinfo {author} {\bibfnamefont {J.~A.}\
			\bibnamefont {Konter}}, \bibinfo {author} {\bibfnamefont {A.}~\bibnamefont
			{Pichlmaier}}, \bibinfo {author} {\bibfnamefont {A.}~\bibnamefont {Wokaun}},
		\bibinfo {author} {\bibfnamefont {K.}~\bibnamefont {Bodek}}, \bibinfo {author} {\bibfnamefont
			{M.}~\bibnamefont {Kasprzak}}, \bibinfo {author} {\bibfnamefont
			{M.}~\bibnamefont {Kuzniak}}, \bibinfo
		{author} {\bibfnamefont {P.}~\bibnamefont {Geltenbort}},\ and\ \bibinfo
		{author} {\bibfnamefont {J.}~\bibnamefont {Zmeskal}},\ }\bibfield  {title}
	{\bibinfo {title} {Measured total cross sections of slow neutrons scattered
			by solid deuterium and implications for ultracold neutron sources},\ }\href
	{https://doi.org/10.1103/PhysRevLett.95.182502} {\bibfield  {journal}
		{\bibinfo  {journal} {Phys. Rev. Lett.}\ }\textbf {\bibinfo {volume} {95}},\
		\bibinfo {pages} {182502} (\bibinfo {year} {2005})}\BibitemShut {NoStop}%
	\bibitem [{\citenamefont {Yu}\ and\ \citenamefont {Golub}(1986)}]{Yu1986}%
	\BibitemOpen
	\bibfield  {author} {\bibinfo {author} {\bibfnamefont {S.}~\bibnamefont {Yu},
			\bibfnamefont {Z.-Ch.and~Malik}}\ and\ \bibinfo {author} {\bibfnamefont
			{R.}~\bibnamefont {Golub}},\ }\bibfield  {title} {\bibinfo {title} {A thin
			film source of ultra-cold neutrons},\ }\href
	{https://doi.org/10.1007/BF01323423} {\bibfield  {journal} {\bibinfo
			{journal} {Z. Phys. B}\ }\textbf {\bibinfo {volume} {62}},\ \bibinfo {pages}
		{137} (\bibinfo {year} {1986})}\BibitemShut {NoStop}%
	\bibitem [{\citenamefont {Becker}\ \emph {et~al.}(2015)\citenamefont {Becker},
		\citenamefont {Bison}, \citenamefont {Blau}, \citenamefont {Chowdhuri},
		\citenamefont {Eikenberg}, \citenamefont {Fertl}, \citenamefont {Kirch},
		\citenamefont {Lauss}, \citenamefont {Perret}, \citenamefont {Reggiani},
		\citenamefont {Ries}, \citenamefont {Schmidt-Wellenburg}, \citenamefont
		{Talanov}, \citenamefont {Wohlmuther},\ and\ \citenamefont
		{Zsigmond}}]{Becker2015}%
	\BibitemOpen
	\bibfield  {author} {\bibinfo {author} {\bibfnamefont {H.}~\bibnamefont
			{Becker}}, \bibinfo {author} {\bibfnamefont {G.}~\bibnamefont {Bison}},
		\bibinfo {author} {\bibfnamefont {B.}~\bibnamefont {Blau}}, \bibinfo {author}
		{\bibfnamefont {Z.}~\bibnamefont {Chowdhuri}}, \bibinfo {author}
		{\bibfnamefont {J.}~\bibnamefont {Eikenberg}}, \bibinfo {author}
		{\bibfnamefont {M.}~\bibnamefont {Fertl}}, \bibinfo {author} {\bibfnamefont
			{K.}~\bibnamefont {Kirch}}, \bibinfo {author} {\bibfnamefont
			{B.}~\bibnamefont {Lauss}}, \bibinfo {author} {\bibfnamefont
			{G.}~\bibnamefont {Perret}}, \bibinfo {author} {\bibfnamefont
			{D.}~\bibnamefont {Reggiani}}, \bibinfo {author} {\bibfnamefont
			{D.}~\bibnamefont {Ries}}, \bibinfo {author} {\bibfnamefont {P.}~\bibnamefont
			{Schmidt-Wellenburg}}, \bibinfo {author} {\bibfnamefont {V.}~\bibnamefont
			{Talanov}}, \bibinfo {author} {\bibfnamefont {M.}~\bibnamefont
			{Wohlmuther}},\ and\ \bibinfo {author} {\bibfnamefont {G.}~\bibnamefont
			{Zsigmond}},\ }\bibfield  {title} {\bibinfo {title} {Neutron production and
			thermal moderation at the {PSI UCN} source},\ }\href
	{https://doi.org/10.1016/j.nima.2014.12.091} {\bibfield  {journal} {\bibinfo
			{journal} {Nucl. Instrum. Methods A}\ }\textbf {\bibinfo {volume} {777}},\
		\bibinfo {pages} {20} (\bibinfo {year} {2015})}\BibitemShut {NoStop}%
	\bibitem [{\citenamefont {Zsigmond}(2018)}]{Zsigmond2018}%
	\BibitemOpen
	\bibfield  {author} {\bibinfo {author} {\bibfnamefont {G.}~\bibnamefont
			{Zsigmond}},\ }\bibfield  {title} {\bibinfo {title} {The {MCUCN} simulation
			code for ultracold neutron physics},\ }\href
	{https://doi.org/10.1016/j.nima.2017.10.065} {\bibfield  {journal} {\bibinfo
			{journal} {Nucl. Instrum. Methods A}\ }\textbf {\bibinfo {volume} {881}},\
		\bibinfo {pages} {16} (\bibinfo {year} {2018})}\BibitemShut {NoStop}%
	\bibitem [{\citenamefont {Goorley}\ \emph {et~al.}(2012)\citenamefont
		{Goorley}, \citenamefont {James}, \citenamefont {Booth}, \citenamefont
		{Brown}, \citenamefont {Bull}, \citenamefont {Cox}, \citenamefont {Durkee},
		\citenamefont {Elson}, \citenamefont {Fensin}, \citenamefont {Forster},
		\citenamefont {Hendricks}, \citenamefont {Hughes}, \citenamefont {Johns},
		\citenamefont {Kiedrowski}, \citenamefont {Martz}, \citenamefont {Mashnik},
		\citenamefont {McKinney}, \citenamefont {Pelowitz}, \citenamefont {Prael},
		\citenamefont {Sweezy}, \citenamefont {Waters}, \citenamefont {Wilcox},\ and\
		\citenamefont {Zukaitis}}]{Goorley2012}%
	\BibitemOpen
	\bibfield  {author} {\bibinfo {author} {\bibfnamefont {T.}~\bibnamefont
			{Goorley}}, \bibinfo {author} {\bibfnamefont {M.}~\bibnamefont {James}},
		\bibinfo {author} {\bibfnamefont {T.}~\bibnamefont {Booth}}, \bibinfo
		{author} {\bibfnamefont {F.}~\bibnamefont {Brown}}, \bibinfo {author}
		{\bibfnamefont {J.}~\bibnamefont {Bull}}, \bibinfo {author} {\bibfnamefont
			{L.~J.}\ \bibnamefont {Cox}}, \bibinfo {author} {\bibfnamefont
			{J.}~\bibnamefont {Durkee}}, \bibinfo {author} {\bibfnamefont
			{J.}~\bibnamefont {Elson}}, \bibinfo {author} {\bibfnamefont
			{M.}~\bibnamefont {Fensin}}, \bibinfo {author} {\bibfnamefont {R.~A.}\
			\bibnamefont {Forster}}, \bibinfo {author} {\bibfnamefont {J.}~\bibnamefont
			{Hendricks}}, \bibinfo {author} {\bibfnamefont {H.~G.}\ \bibnamefont
			{Hughes}}, \bibinfo {author} {\bibfnamefont {R.}~\bibnamefont {Johns}},
		\bibinfo {author} {\bibfnamefont {B.}~\bibnamefont {Kiedrowski}}, \bibinfo
		{author} {\bibfnamefont {R.}~\bibnamefont {Martz}}, \bibinfo {author}
		{\bibfnamefont {S.}~\bibnamefont {Mashnik}}, \bibinfo {author} {\bibfnamefont
			{G.}~\bibnamefont {McKinney}}, \bibinfo {author} {\bibfnamefont
			{D.}~\bibnamefont {Pelowitz}}, \bibinfo {author} {\bibfnamefont
			{R.}~\bibnamefont {Prael}}, \bibinfo {author} {\bibfnamefont
			{J.}~\bibnamefont {Sweezy}}, \bibinfo {author} {\bibfnamefont
			{L.}~\bibnamefont {Waters}}, \bibinfo {author} {\bibfnamefont
			{T.}~\bibnamefont {Wilcox}},\ and\ \bibinfo {author} {\bibfnamefont
			{T.}~\bibnamefont {Zukaitis}},\ }\bibfield  {title} {\bibinfo {title}
		{Initial {MCNP}6 release overview},\ }\href
	{https://doi.org/10.13182/nt11-135} {\bibfield  {journal} {\bibinfo
			{journal} {Nuclear Technology}\ }\textbf {\bibinfo {volume} {180}},\ \bibinfo
		{pages} {298} (\bibinfo {year} {2012})}\BibitemShut {NoStop}%
	\bibitem [{\citenamefont {Atchison}\ \emph {et~al.}(2007)\citenamefont
		{Atchison}, \citenamefont {Blau}, \citenamefont {Bodek}, \citenamefont
		{van~den Brandt}, \citenamefont {Brys}, \citenamefont {Daum}, \citenamefont
		{Fierlinger}, \citenamefont {Frei}, \citenamefont {Geltenbort}, \citenamefont
		{Hautle}, \citenamefont {Henneck}, \citenamefont {Heule}, \citenamefont
		{Holley}, \citenamefont {Kasprzak}, \citenamefont {Kirch}, \citenamefont
		{Knecht}, \citenamefont {Konter}, \citenamefont {Kuzniak}, \citenamefont
		{Liu}, \citenamefont {Morris}, \citenamefont {Pichlmaier}, \citenamefont
		{Plonka}, \citenamefont {Pokotilovski}, \citenamefont {Saunders},
		\citenamefont {Shin}, \citenamefont {Tortorella}, \citenamefont {Wohlmuther},
		\citenamefont {Young}, \citenamefont {Zejma},\ and\ \citenamefont
		{Zsigmond}}]{Atchison2007}%
	\BibitemOpen
	\bibfield  {author} {\bibinfo {author} {\bibfnamefont {F.}~\bibnamefont
			{Atchison}}, \bibinfo {author} {\bibfnamefont {B.}~\bibnamefont {Blau}},
		\bibinfo {author} {\bibfnamefont {K.}~\bibnamefont {Bodek}}, \bibinfo
		{author} {\bibfnamefont {B.}~\bibnamefont {van~den Brandt}}, \bibinfo
		{author} {\bibfnamefont {T.}~\bibnamefont {Brys}}, \bibinfo {author}
		{\bibfnamefont {M.}~\bibnamefont {Daum}}, \bibinfo {author} {\bibfnamefont
			{P.}~\bibnamefont {Fierlinger}}, \bibinfo {author} {\bibfnamefont
			{A.}~\bibnamefont {Frei}}, \bibinfo {author} {\bibfnamefont {P.}~\bibnamefont
			{Geltenbort}}, \bibinfo {author} {\bibfnamefont {P.}~\bibnamefont {Hautle}},
		\bibinfo {author} {\bibfnamefont {R.}~\bibnamefont {Henneck}}, \bibinfo
		{author} {\bibfnamefont {S.}~\bibnamefont {Heule}}, \bibinfo {author}
		{\bibfnamefont {A.}~\bibnamefont {Holley}}, \bibinfo {author} {\bibfnamefont
			{M.}~\bibnamefont {Kasprzak}}, \bibinfo {author} {\bibfnamefont
			{K.}~\bibnamefont {Kirch}}, \bibinfo {author} {\bibfnamefont
			{A.}~\bibnamefont {Knecht}}, \bibinfo {author} {\bibfnamefont {J.~A.}\
			\bibnamefont {Konter}}, \bibinfo {author} {\bibfnamefont {M.}~\bibnamefont
			{Kuzniak}}, \bibinfo {author} {\bibfnamefont {C.-Y.}\ \bibnamefont {Liu}},
		\bibinfo {author} {\bibfnamefont {C.~L.}\ \bibnamefont {Morris}}, \bibinfo
		{author} {\bibfnamefont {A.}~\bibnamefont {Pichlmaier}}, \bibinfo {author}
		{\bibfnamefont {C.}~\bibnamefont {Plonka}}, \bibinfo {author} {\bibfnamefont
			{Y.}~\bibnamefont {Pokotilovski}}, \bibinfo {author} {\bibfnamefont
			{A.}~\bibnamefont {Saunders}}, \bibinfo {author} {\bibfnamefont
			{Y.}~\bibnamefont {Shin}}, \bibinfo {author} {\bibfnamefont {D.}~\bibnamefont
			{Tortorella}}, \bibinfo {author} {\bibfnamefont {M.}~\bibnamefont
			{Wohlmuther}}, \bibinfo {author} {\bibfnamefont {A.~R.}\ \bibnamefont
			{Young}}, \bibinfo {author} {\bibfnamefont {J.}~\bibnamefont {Zejma}},\ and\
		\bibinfo {author} {\bibfnamefont {G.}~\bibnamefont {Zsigmond}},\ }\bibfield
	{title} {\bibinfo {title} {Cold neutron energy dependent production of
			ultracold neutrons in solid deuterium},\ }\href
	{https://doi.org/10.1103/PhysRevLett.99.262502} {\bibfield  {journal}
		{\bibinfo  {journal} {Phys. Rev. Lett.}\ }\textbf {\bibinfo {volume} {99}},\
		\bibinfo {pages} {262502} (\bibinfo {year} {2007})}\BibitemShut {NoStop}%
	\bibitem [{\citenamefont {Frei}\ \emph {et~al.}(2009)\citenamefont {Frei},
		\citenamefont {Gutsmiedl}, \citenamefont {Morkel}, \citenamefont {Müller},
		\citenamefont {Paul}, \citenamefont {Urban}, \citenamefont {Schober},
		\citenamefont {Rols}, \citenamefont {Unruh},\ and\ \citenamefont
		{Hölzel}}]{Frei2009}%
	\BibitemOpen
	\bibfield  {author} {\bibinfo {author} {\bibfnamefont {A.}~\bibnamefont
			{Frei}}, \bibinfo {author} {\bibfnamefont {E.}~\bibnamefont {Gutsmiedl}},
		\bibinfo {author} {\bibfnamefont {C.}~\bibnamefont {Morkel}}, \bibinfo
		{author} {\bibfnamefont {A.~R.}\ \bibnamefont {Müller}}, \bibinfo {author}
		{\bibfnamefont {S.}~\bibnamefont {Paul}}, \bibinfo {author} {\bibfnamefont
			{M.}~\bibnamefont {Urban}}, \bibinfo {author} {\bibfnamefont
			{H.}~\bibnamefont {Schober}}, \bibinfo {author} {\bibfnamefont
			{S.}~\bibnamefont {Rols}}, \bibinfo {author} {\bibfnamefont {T.}~\bibnamefont
			{Unruh}},\ and\ \bibinfo {author} {\bibfnamefont {M.}~\bibnamefont
			{Hölzel}},\ }\bibfield  {title} {\bibinfo {title} {Density of states in solid
			deuterium: Inelastic neutron scattering study},\ }\href
	{https://doi.org/10.1103/physrevb.80.064301} {\bibfield  {journal} {\bibinfo
			{journal} {Physical Review B}\ }\textbf {\bibinfo {volume} {80}},\ \bibinfo
		{pages} {064301} (\bibinfo {year} {2009})}\BibitemShut {NoStop}%
	\bibitem [{\citenamefont {Nielsen}\ and\ \citenamefont
		{M{\o}ller}(1971)}]{Nielsen1971}%
	\BibitemOpen
	\bibfield  {author} {\bibinfo {author} {\bibfnamefont {M.}~\bibnamefont
			{Nielsen}}\ and\ \bibinfo {author} {\bibfnamefont {H.~B.}\ \bibnamefont
			{M{\o}ller}},\ }\bibfield  {title} {\bibinfo {title} {Lattice dynamics of
			solid deuterium by inelastic neutron scattering},\ }\href
	{https://doi.org/10.1103/physrevb.3.4383} {\bibfield  {journal} {\bibinfo
			{journal} {Physical Review B}\ }\textbf {\bibinfo {volume} {3}},\ \bibinfo
		{pages} {4383} (\bibinfo {year} {1971})}\BibitemShut {NoStop}%
	\bibitem [{\citenamefont {Anghel}\ \emph {et~al.}(2008)\citenamefont {Anghel},
		\citenamefont {Blau}, \citenamefont {Daum}, \citenamefont {Kirch},\ and\
		\citenamefont {Grigoriev}}]{Anghel2008}%
	\BibitemOpen
	\bibfield  {author} {\bibinfo {author} {\bibfnamefont {A.}~\bibnamefont
			{Anghel}}, \bibinfo {author} {\bibfnamefont {B.}~\bibnamefont {Blau}},
		\bibinfo {author} {\bibfnamefont {M.}~\bibnamefont {Daum}}, \bibinfo {author}
		{\bibfnamefont {K.}~\bibnamefont {Kirch}},\ and\ \bibinfo {author}
		{\bibfnamefont {S.}~\bibnamefont {Grigoriev}},\ }\bibfield  {title} {\bibinfo
		{title} {Cryogenic system of the {S}wiss ultra-cold neutron source},\ }in\
	\href
	{https://iifiir.org/en/fridoc/cryogenic-system-of-the-swiss-ultracold-neutron-source-25219}
	{\emph {\bibinfo {booktitle} {Proc. of the 10th IIR International Conference
				Cryogenics, 04/21-04/25, 2008, Prague, Czech Republic}}}\ (\bibinfo {year}
	{2008})\BibitemShut {NoStop}%
	\bibitem [{\citenamefont {Lauss}\ and\ \citenamefont {Blau}(2021)}]{Lauss2021}%
	\BibitemOpen
	\bibfield  {author} {\bibinfo {author} {\bibfnamefont {B.}~\bibnamefont
			{Lauss}}\ and\ \bibinfo {author} {\bibfnamefont {B.}~\bibnamefont {Blau}},\
	}\bibfield  {title} {\bibinfo {title} {{UCN}, the ultracold neutron source --
			neutrons for particle physics},\ }\bibfield  {journal} {\bibinfo  {journal}
		{{SciPost} Physics Proceedings}\ }\href
	{https://doi.org/10.21468/scipostphysproc.5.004}
	{10.21468/scipostphysproc.5.004} (\bibinfo {year} {2021})\BibitemShut
	{NoStop}%
	\bibitem [{\citenamefont {Sullivan}\ \emph {et~al.}(1990)\citenamefont
		{Sullivan}, \citenamefont {Zhou},\ and\ \citenamefont
		{Edwards}}]{Sullivan1990}%
	\BibitemOpen
	\bibfield  {author} {\bibinfo {author} {\bibfnamefont {N.}~\bibnamefont
			{Sullivan}}, \bibinfo {author} {\bibfnamefont {D.}~\bibnamefont {Zhou}},\
		and\ \bibinfo {author} {\bibfnamefont {C.}~\bibnamefont {Edwards}},\
	}\bibfield  {title} {\bibinfo {title} {Precise and efficient in situ ortho
			{\textemdash} para-hydrogen converter},\ }\href
	{https://doi.org/10.1016/0011-2275(90)90240-d} {\bibfield  {journal}
		{\bibinfo  {journal} {Cryogenics}\ }\textbf {\bibinfo {volume} {30}},\
		\bibinfo {pages} {734} (\bibinfo {year} {1990})}\BibitemShut {NoStop}%
	\bibitem [{\citenamefont {Bodek}\ \emph {et~al.}(2004)\citenamefont {Bodek},
		\citenamefont {van~den Brandt}, \citenamefont {Brys}, \citenamefont {Daum},
		\citenamefont {Fierlinger}, \citenamefont {Geltenbort}, \citenamefont
		{Giersch}, \citenamefont {Hautle}, \citenamefont {Henneck}, \citenamefont
		{Kasprzak}, \citenamefont {Kirch}, \citenamefont {Konter}, \citenamefont
		{Kuehne}, \citenamefont {Kuzniak}, \citenamefont {Mishima}, \citenamefont
		{Pichlmaier}, \citenamefont {Raetz}, \citenamefont {Serebrov},\ and\
		\citenamefont {Zmeskal}}]{Bodek2004}%
	\BibitemOpen
	\bibfield  {author} {\bibinfo {author} {\bibfnamefont {K.}~\bibnamefont
			{Bodek}}, \bibinfo {author} {\bibfnamefont {B.}~\bibnamefont {van~den
				Brandt}}, \bibinfo {author} {\bibfnamefont {T.}~\bibnamefont {Brys}},
		\bibinfo {author} {\bibfnamefont {M.}~\bibnamefont {Daum}}, \bibinfo {author}
		{\bibfnamefont {P.}~\bibnamefont {Fierlinger}}, \bibinfo {author}
		{\bibfnamefont {P.}~\bibnamefont {Geltenbort}}, \bibinfo {author}
		{\bibfnamefont {M.}~\bibnamefont {Giersch}}, \bibinfo {author} {\bibfnamefont
			{P.}~\bibnamefont {Hautle}}, \bibinfo {author} {\bibfnamefont
			{R.}~\bibnamefont {Henneck}}, \bibinfo {author} {\bibfnamefont
			{M.}~\bibnamefont {Kasprzak}}, \bibinfo {author} {\bibfnamefont
			{K.}~\bibnamefont {Kirch}}, \bibinfo {author} {\bibfnamefont
			{J.}~\bibnamefont {Konter}}, \bibinfo {author} {\bibfnamefont
			{G.}~\bibnamefont {Kuehne}}, \bibinfo {author} {\bibfnamefont
			{M.}~\bibnamefont {Kuzniak}}, \bibinfo {author} {\bibfnamefont
			{K.}~\bibnamefont {Mishima}}, \bibinfo {author} {\bibfnamefont
			{A.}~\bibnamefont {Pichlmaier}}, \bibinfo {author} {\bibfnamefont
			{D.}~\bibnamefont {Raetz}}, \bibinfo {author} {\bibfnamefont
			{A.}~\bibnamefont {Serebrov}},\ and\ \bibinfo {author} {\bibfnamefont
			{J.}~\bibnamefont {Zmeskal}},\ }\bibfield  {title} {\bibinfo {title} {An
			apparatus for the investigation of solid {D2} with respect to ultra-cold
			neutron sources},\ }\href {https://doi.org/10.1016/j.nima.2004.06.157}
	{\bibfield  {journal} {\bibinfo  {journal} {Nucl. Instrum. Methods A}\
		}\textbf {\bibinfo {volume} {533}},\ \bibinfo {pages} {491} (\bibinfo {year}
		{2004})}\BibitemShut {NoStop}%
	\bibitem [{\citenamefont {Hild}(2019)}]{Hild2019}%
	\BibitemOpen
	\bibfield  {author} {\bibinfo {author} {\bibfnamefont {N.}~\bibnamefont
			{Hild}},\ }\emph {\bibinfo {title} {Studies of the Deuterium Used in the PSI
			UCN Source}},\ \href {https://doi.org/10.3929/ETHZ-B-000390393} {Ph.D.
		thesis},\ \bibinfo  {school} {ETHZ Zurich, No.26412} (\bibinfo {year}
	{2019})\BibitemShut {NoStop}%
	\bibitem [{\citenamefont {Collins}\ \emph {et~al.}(1991)\citenamefont
		{Collins}, \citenamefont {Fearon}, \citenamefont {Mapoles}, \citenamefont
		{Tsugawa}, \citenamefont {Souers},\ and\ \citenamefont
		{Fedders}}]{Collins1991}%
	\BibitemOpen
	\bibfield  {author} {\bibinfo {author} {\bibfnamefont {G.~W.}\ \bibnamefont
			{Collins}}, \bibinfo {author} {\bibfnamefont {E.~M.}\ \bibnamefont {Fearon}},
		\bibinfo {author} {\bibfnamefont {E.~R.}\ \bibnamefont {Mapoles}}, \bibinfo
		{author} {\bibfnamefont {R.~T.}\ \bibnamefont {Tsugawa}}, \bibinfo {author}
		{\bibfnamefont {P.~C.}\ \bibnamefont {Souers}},\ and\ \bibinfo {author}
		{\bibfnamefont {P.~A.}\ \bibnamefont {Fedders}},\ }\bibfield  {title}
	{\bibinfo {title} {\textit{J} =1 to \textit{J} =0 ${D}_{2}$ conversion in
			solid {D-T}},\ }\href {https://doi.org/10.1103/PhysRevB.44.6598} {\bibfield
		{journal} {\bibinfo  {journal} {Phys. Rev. B}\ }\textbf {\bibinfo {volume}
			{44}},\ \bibinfo {pages} {6598} (\bibinfo {year} {1991})}\BibitemShut
	{NoStop}%
	\bibitem [{\citenamefont {Atchison}\ \emph {et~al.}(2003)\citenamefont
		{Atchison}, \citenamefont {Beaud}, \citenamefont {Brys}, \citenamefont
		{Daum}, \citenamefont {Fierlinger}, \citenamefont {Henneck}, \citenamefont
		{Hofmann}, \citenamefont {Kirch}, \citenamefont {Kuehne}, \citenamefont
		{Knopp}, \citenamefont {Pichlmaier}, \citenamefont {Serebrov}, \citenamefont
		{Spitzer}, \citenamefont {Wambach}, \citenamefont {Wimmer}, \citenamefont
		{Wokaun}, \citenamefont {Bodek}, \citenamefont {Geltenbort}, \citenamefont
		{Giersch}, \citenamefont {Zmeskal},\ and\ \citenamefont
		{Mishima}}]{Atchison2003}%
	\BibitemOpen
	\bibfield  {author} {\bibinfo {author} {\bibfnamefont {F.}~\bibnamefont
			{Atchison}}, \bibinfo {author} {\bibfnamefont {P.}~\bibnamefont {Beaud}},
		\bibinfo {author} {\bibfnamefont {T.}~\bibnamefont {Brys}}, \bibinfo {author}
		{\bibfnamefont {M.}~\bibnamefont {Daum}}, \bibinfo {author} {\bibfnamefont
			{P.}~\bibnamefont {Fierlinger}}, \bibinfo {author} {\bibfnamefont
			{R.}~\bibnamefont {Henneck}}, \bibinfo {author} {\bibfnamefont
			{T.}~\bibnamefont {Hofmann}}, \bibinfo {author} {\bibfnamefont
			{K.}~\bibnamefont {Kirch}}, \bibinfo {author} {\bibfnamefont
			{G.}~\bibnamefont {Kuehne}}, \bibinfo {author} {\bibfnamefont
			{G.}~\bibnamefont {Knopp}}, \bibinfo {author} {\bibfnamefont
			{A.}~\bibnamefont {Pichlmaier}}, \bibinfo {author} {\bibfnamefont
			{A.}~\bibnamefont {Serebrov}}, \bibinfo {author} {\bibfnamefont
			{H.}~\bibnamefont {Spitzer}}, \bibinfo {author} {\bibfnamefont
			{J.}~\bibnamefont {Wambach}}, \bibinfo {author} {\bibfnamefont
			{J.}~\bibnamefont {Wimmer}}, \bibinfo {author} {\bibfnamefont
			{A.}~\bibnamefont {Wokaun}}, \bibinfo {author} {\bibfnamefont
			{K.}~\bibnamefont {Bodek}}, \bibinfo {author} {\bibfnamefont
			{P.}~\bibnamefont {Geltenbort}}, \bibinfo {author} {\bibfnamefont
			{M.}~\bibnamefont {Giersch}}, \bibinfo {author} {\bibfnamefont
			{J.}~\bibnamefont {Zmeskal}},\ and\ \bibinfo {author} {\bibfnamefont
			{K.}~\bibnamefont {Mishima}},\ }\bibfield  {title} {\bibinfo {title}
		{Ortho-para equilibrium in a liquid d2 neutron moderator under irradiation},\
	}\href {https://doi.org/10.1103/PhysRevB.68.094114} {\bibfield  {journal}
		{\bibinfo  {journal} {Phys. Rev. B}\ }\textbf {\bibinfo {volume} {68}},\
		\bibinfo {pages} {094114} (\bibinfo {year} {2003})}\BibitemShut {NoStop}%
	\bibitem [{\citenamefont {Mishima}\ \emph {et~al.}(2007)\citenamefont
		{Mishima}, \citenamefont {Utsuro}, \citenamefont {Nagai}, \citenamefont
		{Tanaka}, \citenamefont {Kohmoto}, \citenamefont {Fukuda}, \citenamefont
		{Kiyanagi},\ and\ \citenamefont {Ooi}}]{Mishima2007}%
	\BibitemOpen
	\bibfield  {author} {\bibinfo {author} {\bibfnamefont {K.}~\bibnamefont
			{Mishima}}, \bibinfo {author} {\bibfnamefont {M.}~\bibnamefont {Utsuro}},
		\bibinfo {author} {\bibfnamefont {Y.}~\bibnamefont {Nagai}}, \bibinfo
		{author} {\bibfnamefont {M.}~\bibnamefont {Tanaka}}, \bibinfo {author}
		{\bibfnamefont {T.}~\bibnamefont {Kohmoto}}, \bibinfo {author} {\bibfnamefont
			{Y.}~\bibnamefont {Fukuda}}, \bibinfo {author} {\bibfnamefont
			{Y.}~\bibnamefont {Kiyanagi}},\ and\ \bibinfo {author} {\bibfnamefont
			{M.}~\bibnamefont {Ooi}},\ }\bibfield  {title} {\bibinfo {title}
		{Time-differential observation of the ortho-para conversion of liquid d$_2$
			under irradiation},\ }\href {https://doi.org/10.1103/physrevb.75.014112}
	{\bibfield  {journal} {\bibinfo  {journal} {Physical Review B}\ }\textbf
		{\bibinfo {volume} {75}},\ \bibinfo {pages} {014112} (\bibinfo {year}
		{2007})}\BibitemShut {NoStop}%
	\bibitem [{\citenamefont {Wlokka}(2016)}]{Wlokka2016}%
	\BibitemOpen
	\bibfield  {author} {\bibinfo {author} {\bibfnamefont {S.~A.}\ \bibnamefont
			{Wlokka}},\ }\emph {\bibinfo {title} {Aspects of Ultra-Cold Neutron
			Production in Radiation Fields at the FRM II}},\ \href
	{https://mediatum.ub.tum.de/1311462} {Ph.D. thesis},\ \bibinfo  {school}
	{Technische Universität München} (\bibinfo {year} {2016})\BibitemShut
	{NoStop}%
	\bibitem [{\citenamefont {Souers}(1986)}]{Souers1986}%
	\BibitemOpen
	\bibfield  {author} {\bibinfo {author} {\bibfnamefont {P.~C.}\ \bibnamefont
			{Souers}},\ }\href@noop {} {\emph {\bibinfo {title} {Hydrogen properties for
				fusion energy}}}\ (\bibinfo  {publisher} {University of California Press},\
	\bibinfo {address} {Berkeley},\ \bibinfo {year} {1986})\BibitemShut {NoStop}%
	\bibitem [{\citenamefont {Sears}(1992)}]{Sears1992}%
	\BibitemOpen
	\bibfield  {author} {\bibinfo {author} {\bibfnamefont {V.~F.}\ \bibnamefont
			{Sears}},\ }\bibfield  {title} {\bibinfo {title} {Neutron scattering lengths
			and cross sections},\ }\href {https://doi.org/10.1080/10448639208218770}
	{\bibfield  {journal} {\bibinfo  {journal} {Neutron News}\ }\textbf {\bibinfo
			{volume} {3}},\ \bibinfo {pages} {26} (\bibinfo {year} {1992})}\BibitemShut
	{NoStop}%
	\bibitem [{\citenamefont {Altarev}\ \emph
		{et~al.}(2008{\natexlab{a}})\citenamefont {Altarev}, \citenamefont
		{Atchison}, \citenamefont {Daum}, \citenamefont {Frei}, \citenamefont
		{Gutsmiedl}, \citenamefont {Hampel}, \citenamefont {Hartmann}, \citenamefont
		{Heil}, \citenamefont {Knecht}, \citenamefont {Kratz}, \citenamefont {Lauer},
		\citenamefont {Meier}, \citenamefont {Paul}, \citenamefont {Sobolev},\ and\
		\citenamefont {Wiehl}}]{Daum2008}%
	\BibitemOpen
	\bibfield  {author} {\bibinfo {author} {\bibfnamefont {I.}~\bibnamefont
			{Altarev}}, \bibinfo {author} {\bibfnamefont {F.}~\bibnamefont {Atchison}},
		\bibinfo {author} {\bibfnamefont {M.}~\bibnamefont {Daum}}, \bibinfo {author}
		{\bibfnamefont {A.}~\bibnamefont {Frei}}, \bibinfo {author} {\bibfnamefont
			{E.}~\bibnamefont {Gutsmiedl}}, \bibinfo {author} {\bibfnamefont
			{G.}~\bibnamefont {Hampel}}, \bibinfo {author} {\bibfnamefont {F.~J.}\
			\bibnamefont {Hartmann}}, \bibinfo {author} {\bibfnamefont {W.}~\bibnamefont
			{Heil}}, \bibinfo {author} {\bibfnamefont {A.}~\bibnamefont {Knecht}},
		\bibinfo {author} {\bibfnamefont {J.~V.}\ \bibnamefont {Kratz}}, \bibinfo
		{author} {\bibfnamefont {T.}~\bibnamefont {Lauer}}, \bibinfo {author}
		{\bibfnamefont {M.}~\bibnamefont {Meier}}, \bibinfo {author} {\bibfnamefont
			{S.}~\bibnamefont {Paul}}, \bibinfo {author} {\bibfnamefont {Y.}~\bibnamefont
			{Sobolev}},\ and\ \bibinfo {author} {\bibfnamefont {N.}~\bibnamefont
			{Wiehl}},\ }\bibfield  {title} {\bibinfo {title} {Direct experimental
			verification of neutron acceleration by the material optical potential of
			solid $^{2}\mathrm{H}_{2}$},\ }\href
	{https://doi.org/10.1103/PhysRevLett.100.014801} {\bibfield  {journal}
		{\bibinfo  {journal} {Phys. Rev. Lett.}\ }\textbf {\bibinfo {volume} {100}},\
		\bibinfo {pages} {014801} (\bibinfo {year} {2008}{\natexlab{a}})}\BibitemShut
	{NoStop}%
	\bibitem [{\citenamefont {Altarev}\ \emph
		{et~al.}(2008{\natexlab{b}})\citenamefont {Altarev}, \citenamefont {Daum},
		\citenamefont {Frei}, \citenamefont {Gutsmiedl}, \citenamefont {Hampel},
		\citenamefont {Hartmann}, \citenamefont {Heil}, \citenamefont {Knecht},
		\citenamefont {Kratz}, \citenamefont {Lauer}, \citenamefont {Meier},
		\citenamefont {Paul}, \citenamefont {Schmidt}, \citenamefont {Sobolev},
		\citenamefont {Wiehl},\ and\ \citenamefont {Zsigmond}}]{Altarev2008}%
	\BibitemOpen
	\bibfield  {author} {\bibinfo {author} {\bibfnamefont {I.}~\bibnamefont
			{Altarev}}, \bibinfo {author} {\bibfnamefont {M.}~\bibnamefont {Daum}},
		\bibinfo {author} {\bibfnamefont {A.}~\bibnamefont {Frei}}, \bibinfo {author}
		{\bibfnamefont {E.}~\bibnamefont {Gutsmiedl}}, \bibinfo {author}
		{\bibfnamefont {G.}~\bibnamefont {Hampel}}, \bibinfo {author} {\bibfnamefont
			{F.~J.}\ \bibnamefont {Hartmann}}, \bibinfo {author} {\bibfnamefont
			{W.}~\bibnamefont {Heil}}, \bibinfo {author} {\bibfnamefont {A.}~\bibnamefont
			{Knecht}}, \bibinfo {author} {\bibfnamefont {J.~V.}\ \bibnamefont {Kratz}},
		\bibinfo {author} {\bibfnamefont {T.}~\bibnamefont {Lauer}}, \bibinfo
		{author} {\bibfnamefont {M.}~\bibnamefont {Meier}}, \bibinfo {author}
		{\bibfnamefont {S.}~\bibnamefont {Paul}}, \bibinfo {author} {\bibfnamefont
			{U.}~\bibnamefont {Schmidt}}, \bibinfo {author} {\bibfnamefont
			{Y.}~\bibnamefont {Sobolev}}, \bibinfo {author} {\bibfnamefont
			{N.}~\bibnamefont {Wiehl}},\ and\ \bibinfo {author} {\bibfnamefont
			{G.}~\bibnamefont {Zsigmond}},\ }\bibfield  {title} {\bibinfo {title}
		{Neutron velocity distribution from a superthermal solid {2H2} ultracold
			neutron source},\ }\href {https://doi.org/10.1140/epja/i2008-10604-8}
	{\bibfield  {journal} {\bibinfo  {journal} {Eur. Phys. J. A}\ }\textbf
		{\bibinfo {volume} {37}},\ \bibinfo {pages} {9} (\bibinfo {year}
		{2008}{\natexlab{b}})}\BibitemShut {NoStop}%
	\bibitem [{\citenamefont {Anghel}\ \emph {et~al.}(2018)\citenamefont {Anghel},
		\citenamefont {Bailey}, \citenamefont {Bison}, \citenamefont {Blau},
		\citenamefont {Broussard}, \citenamefont {Clayton}, \citenamefont
		{Cude-Woods}, \citenamefont {Daum}, \citenamefont {Hawari}, \citenamefont
		{Hild}, \citenamefont {Huffman}, \citenamefont {Ito}, \citenamefont {Kirch},
		\citenamefont {Korobkina}, \citenamefont {Lauss}, \citenamefont {Leung},
		\citenamefont {Lutz}, \citenamefont {Makela}, \citenamefont {Medlin},
		\citenamefont {Morris}, \citenamefont {Pattie}, \citenamefont {Ries},
		\citenamefont {Saunders}, \citenamefont {Schmidt-Wellenburg}, \citenamefont
		{Talanov}, \citenamefont {Young}, \citenamefont {Wehring}, \citenamefont
		{White}, \citenamefont {Wohlmuther},\ and\ \citenamefont
		{Zsigmond}}]{Anghel2018}%
	\BibitemOpen
	\bibfield  {author} {\bibinfo {author} {\bibfnamefont {A.}~\bibnamefont
			{Anghel}}, \bibinfo {author} {\bibfnamefont {T.~L.}\ \bibnamefont {Bailey}},
		\bibinfo {author} {\bibfnamefont {G.}~\bibnamefont {Bison}}, \bibinfo
		{author} {\bibfnamefont {B.}~\bibnamefont {Blau}}, \bibinfo {author}
		{\bibfnamefont {L.~J.}\ \bibnamefont {Broussard}}, \bibinfo {author}
		{\bibfnamefont {S.~M.}\ \bibnamefont {Clayton}}, \bibinfo {author}
		{\bibfnamefont {C.}~\bibnamefont {Cude-Woods}}, \bibinfo {author}
		{\bibfnamefont {M.}~\bibnamefont {Daum}}, \bibinfo {author} {\bibfnamefont
			{A.}~\bibnamefont {Hawari}}, \bibinfo {author} {\bibfnamefont
			{N.}~\bibnamefont {Hild}}, \bibinfo {author} {\bibfnamefont {P.}~\bibnamefont
			{Huffman}}, \bibinfo {author} {\bibfnamefont {T.~M.}\ \bibnamefont {Ito}},
		\bibinfo {author} {\bibfnamefont {K.}~\bibnamefont {Kirch}}, \bibinfo
		{author} {\bibfnamefont {E.}~\bibnamefont {Korobkina}}, \bibinfo {author}
		{\bibfnamefont {B.}~\bibnamefont {Lauss}}, \bibinfo {author} {\bibfnamefont
			{K.}~\bibnamefont {Leung}}, \bibinfo {author} {\bibfnamefont {E.~M.}\
			\bibnamefont {Lutz}}, \bibinfo {author} {\bibfnamefont {M.}~\bibnamefont
			{Makela}}, \bibinfo {author} {\bibfnamefont {G.}~\bibnamefont {Medlin}},
		\bibinfo {author} {\bibfnamefont {C.~L.}\ \bibnamefont {Morris}}, \bibinfo
		{author} {\bibfnamefont {R.~W.}\ \bibnamefont {Pattie}}, \bibinfo {author}
		{\bibfnamefont {D.}~\bibnamefont {Ries}}, \bibinfo {author} {\bibfnamefont
			{A.}~\bibnamefont {Saunders}}, \bibinfo {author} {\bibfnamefont
			{P.}~\bibnamefont {Schmidt-Wellenburg}}, \bibinfo {author} {\bibfnamefont
			{V.}~\bibnamefont {Talanov}}, \bibinfo {author} {\bibfnamefont {A.~R.}\
			\bibnamefont {Young}}, \bibinfo {author} {\bibfnamefont {B.}~\bibnamefont
			{Wehring}}, \bibinfo {author} {\bibfnamefont {C.}~\bibnamefont {White}},
		\bibinfo {author} {\bibfnamefont {M.}~\bibnamefont {Wohlmuther}},\ and\
		\bibinfo {author} {\bibfnamefont {G.}~\bibnamefont {Zsigmond}},\ }\bibfield
	{title} {\bibinfo {title} {Solid deuterium surface degradation at ultracold
			neutron sources},\ }\href {https://doi.org/10.1140/epja/i2018-12594-2}
	{\bibfield  {journal} {\bibinfo  {journal} {The European Physical Journal A}\
		}\textbf {\bibinfo {volume} {54}},\ \bibinfo {pages} {148} (\bibinfo {year}
		{2018})}\BibitemShut {NoStop}%
	\bibitem [{\citenamefont {Ries}(2016)}]{Ries2016}%
	\BibitemOpen
	\bibfield  {author} {\bibinfo {author} {\bibfnamefont {D.}~\bibnamefont
			{Ries}},\ }\emph {\bibinfo {title} {{Characterisation and Optimisation of the
				Source for Ultracold Neutrons at the Paul Scherrer Institute}}},\ \href
	{https://doi.org/10.3929/ethz-a-010795050} {Ph.D. thesis},\ \bibinfo
	{school} {ETH Z{\"u}rich, No.23671} (\bibinfo {year} {2016})\BibitemShut
	{NoStop}%
	\bibitem [{\citenamefont {Atchison}\ \emph {et~al.}(2009)\citenamefont
		{Atchison}, \citenamefont {Blau}, \citenamefont {Bollhalder}, \citenamefont
		{Daum}, \citenamefont {Fierlinger}, \citenamefont {Geltenbort}, \citenamefont
		{Hampel}, \citenamefont {Kasprzak}, \citenamefont {Kirch}, \citenamefont
		{Köchli}, \citenamefont {Kuczewski}, \citenamefont {Leber}, \citenamefont
		{Locher}, \citenamefont {Meier}, \citenamefont {Ochse}, \citenamefont
		{Pichlmaier}, \citenamefont {Plonka}, \citenamefont {Reiser}, \citenamefont
		{Ulrich}, \citenamefont {Wang}, \citenamefont {Wiehl}, \citenamefont
		{Zimmer},\ and\ \citenamefont {Zsigmond}}]{Atchison2009}%
	\BibitemOpen
	\bibfield  {author} {\bibinfo {author} {\bibfnamefont {F.}~\bibnamefont
			{Atchison}}, \bibinfo {author} {\bibfnamefont {B.}~\bibnamefont {Blau}},
		\bibinfo {author} {\bibfnamefont {A.}~\bibnamefont {Bollhalder}}, \bibinfo
		{author} {\bibfnamefont {M.}~\bibnamefont {Daum}}, \bibinfo {author}
		{\bibfnamefont {P.}~\bibnamefont {Fierlinger}}, \bibinfo {author}
		{\bibfnamefont {P.}~\bibnamefont {Geltenbort}}, \bibinfo {author}
		{\bibfnamefont {G.}~\bibnamefont {Hampel}}, \bibinfo {author} {\bibfnamefont
			{M.}~\bibnamefont {Kasprzak}}, \bibinfo {author} {\bibfnamefont
			{K.}~\bibnamefont {Kirch}}, \bibinfo {author} {\bibfnamefont
			{S.}~\bibnamefont {Köchli}}, \bibinfo {author} {\bibfnamefont
			{B.}~\bibnamefont {Kuczewski}}, \bibinfo {author} {\bibfnamefont
			{H.}~\bibnamefont {Leber}}, \bibinfo {author} {\bibfnamefont
			{M.}~\bibnamefont {Locher}}, \bibinfo {author} {\bibfnamefont
			{M.}~\bibnamefont {Meier}}, \bibinfo {author} {\bibfnamefont
			{S.}~\bibnamefont {Ochse}}, \bibinfo {author} {\bibfnamefont
			{A.}~\bibnamefont {Pichlmaier}}, \bibinfo {author} {\bibfnamefont
			{C.}~\bibnamefont {Plonka}}, \bibinfo {author} {\bibfnamefont
			{R.}~\bibnamefont {Reiser}}, \bibinfo {author} {\bibfnamefont
			{J.}~\bibnamefont {Ulrich}}, \bibinfo {author} {\bibfnamefont
			{X.}~\bibnamefont {Wang}}, \bibinfo {author} {\bibfnamefont {N.}~\bibnamefont
			{Wiehl}}, \bibinfo {author} {\bibfnamefont {O.}~\bibnamefont {Zimmer}},\ and\
		\bibinfo {author} {\bibfnamefont {G.}~\bibnamefont {Zsigmond}},\ }\bibfield
	{title} {\bibinfo {title} {Transmission of very slow neutrons through
			material foils and its influence on the design of ultracold neutron
			sources},\ }\href {https://doi.org/10.1016/j.nima.2009.06.047} {\bibfield
		{journal} {\bibinfo  {journal} {Nuclear Instruments and Methods in Physics
				Research Section A: Accelerators, Spectrometers, Detectors and Associated
				Equipment}\ }\textbf {\bibinfo {volume} {608}},\ \bibinfo {pages} {144}
		(\bibinfo {year} {2009})}\BibitemShut {NoStop}%
\end{thebibliography}
%

\end{document}